\documentclass[useAMS,usenatbib,usegraphicx,usedcolumn]{mn2e}
\usepackage{fix2col}

\newcommand{\CIV}{\mbox{C\,{\sc iv}}}
\newcommand{\CII}{\mbox{C\,{\sc ii}}}
\newcommand{\CIII}{\mbox{C\,{\sc iii}}}

\newcommand{\SiIII}{\mbox{Si\,{\sc iii}}}
\newcommand{\SiIV}{\mbox{Si\,{\sc iv}}}
\newcommand{\AlIII}{\mbox{Al\,{\sc iii}}}
\newcommand{\OIII}{\mbox{O\,{\sc iii}}}
\newcommand{\NV}{\mbox{N\,{\sc v}}}
\newcommand{\NIV}{\mbox{N\,{\sc iv}}}
\newcommand{\NIII}{\mbox{N\,{\sc iii}}}
\newcommand{\NeVIII}{\mbox{Ne\,{\sc viii}}}
\newcommand{\OVI}{\mbox{O\,{\sc vi}}}
\newcommand{\OIV}{\mbox{O\,{\sc iv}}}
\newcommand{\OI}{\mbox{O\,{\sc i}}}
\newcommand{\MgII}{\mbox{Mg\,{\sc ii}}}

\newcommand{\HII}{\mbox{H\,{\sc ii}}}

\newcommand{\HeII}{\mbox{He\,{\sc ii}}}
\newcommand{\Ho}{\mbox{H$^0$}}
\newcommand{\Lya}{\mbox{Ly\,$\rm\alpha$}}
\newcommand{\Hb}{\mbox{H\,$\rm\beta$}}
\newcommand{\FeII}{\mbox{Fe\,{\sc ii}}}
\newcommand{\FeXXIII}{\mbox{Fe\,{\sc xxiii}}}
\newcommand{\FeXXII}{\mbox{Fe\,{\sc xxii}}}
\newcommand{\FeXX}{\mbox{Fe\,{\sc xx}}}
\newcommand{\kms}{km~s$^{-1}$}

\newcommand{\ergs}{erg~s$^{-1}$}
\newcommand{\cmmt}{cm$^{-2}$}
\newcommand{\cmt}{cm$^{-3}$}

\newcommand{\aox}{\mbox{$\alpha_{\rm ox}$}}
\newcommand{\aion}{\mbox{$\alpha_{\rm ion}$}}

\newcommand{\NH}{\mbox{$\Sigma_{\rm H}$}}

\newcommand{\Pl}{\mbox{$P_{\rm line}$}}

\newcommand{\hnu}{\mbox{$\langle h\nu\rangle$}}
\newcommand{\Tcomp}{\mbox{$T_{\rm C}$}}
\newcommand{\nH}{\mbox{$n_{\rm H}$}}
\newcommand{\nHi}{\mbox{$n_{{\rm H,i}}$}}
\newcommand{\nHb}{\mbox{$n_{{\rm H,b}}$}}
\newcommand{\ri}{\mbox{$r_{\rm i}$}}
\newcommand{\rb}{\mbox{$r_{\rm b}$}}

\newcommand{\dd}{\textrm{d}}

\newcommand{\tdyn}{\mbox{$t_{\rm dyn}$}}

\title[RPC. II. Application to the BLR]{Radiation pressure confinement -- II. Application to the broad line region in active galactic nuclei}

\author[A.~Baskin, A.~Laor and J.~Stern]
{Alexei Baskin,\thanks{E-mail: alexei@physics.technion.ac.il} 
Ari Laor and Jonathan Stern\\
Physics Department, Technion -- Israel Institute of Technology, Haifa~32000, Israel}

\begin{document}
\date{}
\pagerange{\pageref{firstpage}--\pageref{lastpage}} \pubyear{2013}
\maketitle
\label{firstpage}

\begin{abstract}
Active galactic nuclei (AGN) are characterized by similar broad emission lines properties at all luminosities ($10^{39}$--$10^{47}$~\ergs). What produces this  similarity over a vast range of $10^8$ in luminosity? Photoionization is inevitably associated with momentum transfer to the photoionized gas. Yet, most of the photoionized gas in the Broad Line Region (BLR) follows Keplerian orbits, which suggests that the BLR originates from gas with a large enough column for gravity to dominate. The photoionized surface layer of the gas must develop a pressure gradient due to the incident radiation force. We present solutions for the structure of such a hydrostatic photoionized gas layer in the BLR. The gas is stratified, with a low-density highly-ionized surface layer, a density rise inwards, and a uniform-density cooler inner region, where the gas pressure reaches the incident radiation pressure. This radiation pressure confinement (RPC) of the photoionized layer leads to a universal ionization parameter $U\sim 0.1$ in the inner photoionized layer, independent of luminosity and distance. Thus, RPC appears to explain the universality of the BLR properties in AGN.  We present predictions for the BLR emission per unit covering factor, as a function of distance from the ionizing source, for a range of ionizing continuum slopes and gas metallicity. The predicted mean strength of most lines (excluding \Hb), and their different average-emission radii, are consistent with the available observations.

\end{abstract}
\begin{keywords}
galaxies: active -- quasars: emission lines -- quasars: general.
\end{keywords}

\section{Introduction}\label{sec:intro}

Type 1 active galactic nuclei (AGN) are defined by the presence of broad emission lines, produced by photoionization of the central continuum source \citep{osterbrock_ferland06}. The relative strength of the lines depends on the photoionized gas electron density $n_e$, the ionizing photon density $n_{\gamma}$ and mean energy \hnu, the gas metallicity $Z$, column density \NH, and internal velocity dispersion. Yet, despite the large effect of these free parameters on the relative line strengths, in particular the effects of $n_e$ and the ionization parameter $U\equiv n_{\gamma}/n_e$ (e.g.\ \citealt{korista_etal97}),  the observed properties of the broad lines show a rather small dispersion (e.g.\ \citealt{shen_etal11}). A particularly striking property is the similar relative line strength over the vast range of $10^8$ in luminosity, from the least luminous type 1 AGN at $10^{39}$~\ergs\ (e.g.\ NGC 4395, \citealt{kraemer_etal99}) to the most luminous quasars at $10^{47}$~\ergs\ \citep{croom_etal02, dietrich_etal02}. 

The first photoionization models indicated that the strength of observed emission lines, which are produced by ions that cover a large  range of ionization states, can be produced by a single uniform-density cloud model with $n_e\sim 10^{9.5}$~cm$^{-3}$  and $U\sim 0.01$. The high ionization lines originate from the cloud surface layer, and the low ionization lines from the inner partially ionized  region deep within the cloud \citep{davidson_netzer79}. Followup reverberation mapping (RM) studies implied a smaller BLR radius and thus higher typical values of $n_e\sim 10^{10}$~cm$^{-3}$  and $U\sim 0.1$ (e.g.\ \citealt{rees_etal89}). Some line ratios suggested that part of the BLR may extend up to $n_e\sim 10^{12}$~cm$^{-3}$ (e.g.\ \citealt{baldwin_etal96}) and $U\sim 10$ (e.g.\ \citealt{hamann_etal98}).

Another remarkable BLR property is the uniformity in the values of $n_{\gamma}$. This parameter is given by
\[
n_\gamma=L_{\rm ion}/4\pi r_{\rm BLR}^2 \hnu c,
\]
where $r_{\rm BLR}$ is the radius of the BLR, and $L_{\rm ion}$ is the ionizing luminosity. Reverberation mappings yield
\[
 r_{\rm BLR}\propto L^{0.5},
\]
where $L$ is the near UV luminosity ($\sim$0.5 of $L_{\rm ion}$). The relation extends over a range of $10^7$ in $L$ \citep{kaspi_etal07}, which implies a fixed $n_{\gamma}$, independent of $L$. This relation is well explained by dust sublimation, which sets the outer radius of the BLR \citep{netzer_laor93, suganuma_etal06}, and probably also the inner radius, if the BLR is a dust driven failed wind from the accretion disc surface \citep{czerny_hryniewicz11}. 

So, although we now have a good understanding of what sets $n_{\gamma}$, i.e.\ $n_e U$,  it is not clear why $U\sim 0.1$ is generally preferred, independent of luminosity.  Why are there essentially no AGN where the BLR is dominated by say $U\sim 10^{-3}$ gas? \citet{baldwin_etal95} suggested this is just a selection effect, where there is a broad chaotic distribution with no preferred values of $U$ and $n_e$, and the emission is dominated by the LOC -- Locally Optimally emitting Clouds. However, as noted by \citet{baldwin_etal95}, the LOC clouds must maintain a significant covering factor ($\Omega_{\rm BLR}$) of the ionizing continuum source in order to produce the observed line strength. This constraint is inconsistent with the assumed broad distribution of $U$ values, as clouds with the preferred $U$ values inevitably constitute only a small fraction of the total distribution, and thus of the total covering factor, while observations indicate that the BLR clouds, mostly of $U=0.1$, have $\Omega_{\rm BLR}\simeq 0.3$ \citep{korista_etal97,maiolino_elal01, ruff_etal12}. Another way to phrase this problem is that if the BLR clouds have a power law distribution of $\Omega_{\rm BLR}$ versus $U$, then observations require a specific distribution.  \citet{baldwin_etal95} suggested the observed uniformity is ``simple averaging, not a hidden hand of unknown physics, is at work''. Below we point out the physics which may be at work in setting a universal $U\sim 0.1$.

Another open question is what confines the BLR gas? A two phased photoionized medium was ruled out by \citet{mathews_ferland87}, as the hot phase is too cold ($T<10^7$~K) and therefore dense ($n_e>10^7$~\cmt) to remain invisible in emission. Other confining mechanisms, such as magnetic confinement \citep*{rees87, emmering_etal92}, or a hotter confining medium heated by other mechanisms such as cosmic rays \citep*{begelman_etal91}, are possible additional options. Alternatively, the BLR gas may not need confinement at all, and may be a transient thermal instability in a hot medium (\citealt{krolik88}; cf.\ \citealt{mathews_doane90}), or a transient shock in a hot wind \citep{perry_dyson85}.

Another remarkable coincidence is that the ionizing radiation pressure at the BLR ($P_{\rm rad}$), is similar to the gas pressure ($P_{\rm gas}$), deep enough in the photoionized layer where most of $L_{\rm ion}$ is absorbed. The ionizing radiation pressure is
\[
P_{\rm rad}=L_{\rm ion}/4\pi r_{\rm BLR}^2c
\]
i.e.\ $n_{\gamma}\hnu$, and
\[
 P_{\rm gas}=2n_ekT.
\]
The similarity of $P_{\rm rad}$ and $P_{\rm gas}$ can be seen from the relation
\[
P_{\rm rad}/P_{\rm gas}=n_{\gamma}\hnu/2n_ekT =U\times \hnu/2kT,
\]
adopting $U\sim 0.1$,  $\hnu\sim 30$~eV, and $kT\sim 1$~eV, which yields $P_{\rm rad}/P_{\rm gas}\sim 1$. This similarity was first pointed out by \citet{davidson72}, who however derived $P_{\rm rad}/P_{\rm gas}\sim 1/20$. \citet{shields78} suggested that the photoionized BLR gas may be confined by radiation pressure, but the required
\[
P_{\rm rad}/P_{\rm gas}\sim 1
\]
was inconsistent with the ratio derived from photoionization modelling at those pre-RM days. Direct measurements of $r_{\rm BLR}$ using RM yielded a significantly smaller value than previously estimated from photoionization modelling, thus a higher $P_{\rm rad}$ and $P_{\rm rad}/P_{\rm gas}\sim 1$, as noted above. This relation strongly suggests that Radiation Pressure Confinement (RPC) is the mechanism which sets $P_{\rm gas}$ in the BLR. RPC yields
\[
U=2kT/\hnu.
\]
In photoionized gas $T\sim 10^4$~K is a typical value, and in AGN the mean spectral energy distribution (SED), i.e.\ \hnu, is only weakly luminosity dependent. Thus, RPC implies that $U$ should be a rather universal quantity in AGN, with a value of $\sim 0.1$, as observed. 

To summarize, the observed universal BLR emission line ratios may result from two effects: (i) dust sublimation, which sets $n_e U$, and (ii) RPC, which sets $U$.  The two effects then lead to the similar $n_e$ and $U$ values observed over a range of $10^8$ in luminosity. Finer effects in the emission properties, like the Baldwin relation \citep{baldwin77b} and the eigenvector 1 set of correlations \citep{boroson_green92}, may be produced by systematic trends in the SED shape and metallicity with luminosity (e.g.\ \citealt{korista_etal98, shemmer_etal04}).

Radiation pressure was invoked in earlier studies of the BLR as a mechanism to produce a radial outflow velocity field (e.g.\ \citealt*{blumenthal_mathews75, capriotti_etal81}). However, the generally similar response time of the red and blue wings of the broad emission lines excluded a pure radial flow (e.g.\ \citealt{maoz_etal91}). Furthermore, the similar black hole mass ($M_{\rm BH}$) bulge luminosity relation in nearby galaxies \citep{magorrian_etal98} and in AGN, based on the BLR \citep{laor98}, suggested the BLR velocity field is dominated by gravity, which again argues against a radial flow. Thus, on average the radial velocity at the BLR is zero, and the gas can be considered as approximately radially static. However, photoionization is inevitably associated with momentum transfer from the incident radiation to the gas, and therefore if the photoionized gas is not accelerated radially, it must be subject to an opposing gradient in $P_{\rm gas}$, which balances $P_{\rm rad}$ and allows a hydrostatic solution. In contrast with the uniform density photoionized slab solution (e.g.\ \citealt{korista_etal97}), a solution which includes $P_{\rm rad}$ must have $P_{\rm gas}$ which increases from  $P_{\rm gas,i}$ at the illuminated face of the gas, to $P_{\rm gas,i}+P_{\rm rad}$ deep enough in the gas, where all the ionizing radiation is absorbed. If $P_{\rm rad}\gg P_{\rm gas,i}$, then the gas should also display a large gradient in $n_e$, and thus a larger gradient in ionization states, compared to a uniform $n_e$ solution.

The effect of radiation pressure on photoionized gas in AGN was first explored by \citet{binette_etal97}, in a study of the narrow line emission from the Circinus galaxy. This effect was further explored by \citet{dopita_etal02} (see also \citealt*{groves_etal04}), in a general study of photoionized dusty gas in the Narrow Line Region (NLR) of AGN, where they demonstrated that radiation pressure explains the small range in the observed narrow line properties. A hydrostatic solution of photoionized gas with radiation pressure was presented by \citet{rozanska_etal06} for X-ray warm absorbers in AGN. An analytic hydrostatic solution for dusty \HII\ regions was recently presented by \citet{draine11a}, and a numerical solution was explored by \citet{yeh_etal2013}. 

Here we present a hydrostatic solution for gas in the BLR. In Section~\ref{sec:theory}, we first provide a simplified analytic solution for the density profile inside an RPC photoionized slab, and then discuss the numerical methods used. Section~\ref{sec:results} presents the resulting internal structure of the photoionized gas, and its dependence on the boundary conditions and distance from the ionizing source. We also present the resulting line strength as a function of distance, ionizing SED and metallicity. The results are discussed in Section~\ref{sec:discuss}, and the main conclusions are provided in Section~\ref{sec:conclus}. In a companion paper (\citealt{stern_etal13}, hereafter Paper I), we expand the work of \citet{dopita_etal02} and \citet{groves_etal04}, and study the effect of RPC at all radii beyond the sublimation radius.

\section{Radiation Pressure Confinement}\label{sec:theory}
To explore the net effect of continuum radiation pressure on the radial structure of the BLR gas, we analyze the gas element in its rotating frame, and implicitly assume that the force which determines the radial structure is set only by the incident radiation. This assumption is valid under the following conditions.
\begin{enumerate}
\item The total gas column density is large enough ($\ga10^{24}$~\cmmt) for the black hole gravity to dominate the radiation force.
\item The gas is on circular orbits maintained by the black hole gravity.
\item The gas forms an azimuthally symmetric structure, so a shear in the tangential velocity with radius will not affect the gas radial structure.
\end{enumerate}
The effects of deviations from the above assumptions are briefly discussed in Section~\ref{sec:discuss}.

\subsection{A simplified analytic solution}\label{sec:th_qual}
\subsubsection{The gas pressure structure}
The radiation force on a thin layer of gas at a given $r$ is the momentum deposited by radiation per unit area per unit time, which is
\begin{equation}
f_{\rm rad}=\frac{L_{\rm ion}}{4\pi r^2c} e^{-\tau(r)}\dd\tau ,
\end{equation}
where $\dd\tau$ is the flux weighted mean optical depth of the thin layer, and 
\begin{equation}
\tau(r)=\int_{r_{\rm i}}^r \dd\tau ,
\end{equation}
and $\ri$ is the position of the illuminated face of the slab of gas. For simplicity we assume below that $(r-\ri)/\ri\ll 1$, so one can ignore the geometric dilution of the radiative flux
\begin{equation}
F_{\rm rad}=L_{\rm ion}/4\pi r^2, 
\end{equation}
as it propagates through the slab. Therefore, below we assume $F_{\rm rad}$ is a constant. The relevant luminosity may be somewhat larger than $L_{\rm ion}$, if non ionizing luminosity is also absorbed. For example, in fully ionized gas, where electron scattering dominates, the relevant luminosity is the bolometric luminosity, where $L_{\rm bol}\simeq 2L_{\rm ion}$. In a hydrostatic solution, the radiative force is balanced by the gradient in the gas pressure
\begin{equation}
\dd P_{\rm gas}(r)=\frac{F_{\rm rad}}{c}e^{-\tau(r)}\dd\tau. \label{eq:dPdtau}
\end{equation}
Since $\dd\tau=\alpha \dd r$, where $\alpha$ is the flux weighted mean absorption coefficient and $\dd r$ is the thickness of the layer, we get
\begin{equation}
\frac{\dd P_{\rm gas}(r)}{\dd r}=\frac{F_{\rm rad}}{c}e^{-\tau(r)}\alpha . \label{eq:dP_dr}
\end{equation}
The solution for $P_{\rm gas}(r)$ is given by integrating the above equation. Deep enough in the slab, where $\tau(r)\gg 1$, the integral yields 
\begin{equation}
P_{\rm gas}=\frac{F_{\rm rad}}{c}+P_{\rm gas,i} . \label{eq:P_gas}
\end{equation}
where $P_{\rm gas,i}$ is the gas pressure at the illuminated face, given by the ambient pressure (e.g.\ a hot dilute gas). The solution is independent of the nature of $\alpha$, which just sets the physical scale required to obtain $\tau(r)\gg 1$. Since
\begin{equation}
P_{\rm rad}=F_{\rm rad}/c
\end{equation}
the solution is simply
\begin{equation}
P_{\rm gas}=P_{\rm rad}+P_{\rm gas,i}.\label{eq:P_gas_vs_Prad}
\end{equation}
If the ambient pressure is negligible, i.e.\ $P_{\rm gas,i}\ll P_{\rm rad}$, then at $\tau(r)\gg 1$ we get $P_{\rm gas}=P_{\rm rad}$, i.e.\ the gas pressure deep within the photoionized layer is uniform and is set by the incident radiation pressure, independent of the ambient pressure. Thus, the photoionized gas layer is confined from the illuminated side by the incident radiation pressure, while on the back side it is confined by a thick and static neutral medium.

\subsubsection{The ionization parameter structure}
As noted above, since
\begin{equation}
P_{\rm rad}/P_{\rm gas}=n_{\gamma}\hnu/2n_ekT, 
\end{equation}
RPC yields 
\begin{equation}
n_{\gamma}/n_e=2kT/\hnu,
\end{equation}
deep enough where most of the ionizing radiation is absorbed. This corresponds to $U\sim 0.05$ for photoionized gas at $T\sim 10^4$~K. The value of $U$ increases towards the illuminated surface, and the characteristic value will be $U\sim 0.1$, the value where half of the  ionizing radiation is absorbed. At the illuminated face, $U$ is set by the boundary condition, $U_{\rm i}=n_{\gamma}/n_{e,{\rm i}}$. But, if $P_{\rm gas,i}\ll P_{\rm rad}$, then deep enough $U$ is set only by $P_{\rm rad}$, and is independent of $U_{\rm i}$. The emission structure will also be independent of the boundary values. For example, close to the surface layer, where say only 1 per cent of $F_{\rm rad}$ is absorbed, we necessarily get $P_{\rm gas}=0.01P_{\rm rad}$. This corresponds to 
\begin{equation}
U=5T/10^4,
\end{equation}
or $U\sim 100$ once $T$ is calculated self consistently (see below). This implies that a fixed fraction of about $\sim$1 per cent of the ionizing continuum is reprocessed into line emission in $U\sim 100$ gas. Similarly $\sim$0.1 per cent will be emitted by $U\sim 1000$ gas, $\sim$10 per cent by $U\sim 1-10$ gas, and $\sim$50 per cent by $U=0.05-0.1$ gas. Thus, one does not need to invoke a population of `clouds' with a range of $U$, but rather a single ionized layer produces lines from gas with a large range of $U$,
with well determined relative strengths, set by the RPC solution.

\subsubsection{The density structure}
Specific solutions for $n_e(r)$ based on photoionization calculations are presented below. A simple analytic solution can be obtained for a hot scattering-dominated gas, where  $\alpha=n_e\sigma_{\rm es}$,  and $\sigma_{\rm es}$ is the electron scattering cross section. This condition applies for a low $n_e$ gas, where $U>10^3$ and the gas is fully ionized. The gas will be at the Compton temperature $\Tcomp $, and thus isothermal, which simplifies the problem. We make a further simplification that $\tau(r)<1$, so $e^{-\tau(r)}\sim 1$. The derived equation is then 
\begin{equation}
2k\Tcomp \frac{\dd n_e(r)}{\dd r}=\frac{F_{\rm rad}}{c}n_e\sigma_{\rm es}, 
\end{equation}
which gives 
\begin{equation}
n_e(r)=n_{e,{\rm i}}\exp\left(\frac{r-r_{\rm i}}{l_{\rm pr}}\right),
\end{equation}
where  
\begin{equation}
l_{\rm pr}=2k\Tcomp c/F_{\rm rad}\sigma_{\rm es}. \label{eq:r_pr}
\end{equation}
Thus, there is an exponential rise in $n_e$ on a length scale of $l_{\rm pr}$ (a derivation of $n(r)$, without neglecting the geometrical dilution of $F_{\rm rad}$, is presented in Appendix~\ref{sec:large_d}). Within a few $l_{\rm pr}$ the value of $n_e$ will be high enough to inevitably lead to $U<10^3$, at which point the gas becomes cooler and only partially ionized, leading to a sharp increase in the absorption opacity. Both effects, $T\ll \Tcomp $ and $\sigma\gg \sigma_{\rm es}$ will produce a sharp rise in $\dd n_e/\dd r$.\footnote{For $U<10^3$, where $T<\Tcomp$, the gas $\sigma$ is a function of $T$, and there is no analytic solution for $n_e(r)$. This is in contrast with Paper I, where the opacity down to $U\approx10^{-2}$ is dominated by dust, which has a fixed $\sigma$.}  The uniform pressure region, i.e.\ $\tau(r)\gg 1$, is therefore reached within a few $l_{\rm pr}$ at most, even for an illuminated face with a very low density.

To estimate the value of $l_{\rm pr}$ we need the values of $F_{\rm rad}$ and $\Tcomp $ at the BLR. The value of $F_{\rm rad}$ is derived from the observed relation
\begin{equation}
r_{\rm BLR}\approx 0.1L^{0.5}_{46}\mbox{~pc},
\end{equation}
where $L_{46}=L/10^{46}$~\ergs\ and $L$ is the bolometric luminosity (\citealt{kaspi_etal05}; assuming $L=3L_{1350}$, where $L_{1350}$ is the luminosity at 1350~\AA). This implies 
\begin{equation}\label{eq:eval_F_tot}
 F_{\rm rad}= L/4\pi r_{\rm BLR}^2 \approx 10^{10}{\rm erg~cm}^{-2}~{\rm s}^{-1}. 
\end{equation}
Photoionization calculations yield $\Tcomp \sim 3\times 10^6$~K (see below), which gives $l_{\rm pr}\approx 4\times 10^{15}$~cm, independent of the AGN luminosity. This size becomes larger than $r_{\rm BLR}$ for $L<10^{42}$~\ergs. In such low luminosity AGN, RPC can still confine the BLR gas if $U<10^3$ at the illuminated face, as the gas is not fully ionized, leading to $\sigma_{\rm abs}/\sigma_{\rm es}\sim 10-1000$ and $T<10^6$~K, which reduces $l_{\rm pr}$ by a few orders of magnitude, allowing RPC to work in the lowest luminosity AGN. 

Since
\begin{equation}
 P_{\rm rad}=F_{\rm rad}/c\approx 0.3\mbox{~erg~cm$^{-3}$}
\end{equation}
at the BLR, $P_{\rm gas}=P_{\rm rad}$ implies
\begin{equation}
 n_eT\approx 10^{15}\mbox{~\cmt~K},
\end{equation}
or $n_e\approx 10^{11}$~\cmt\ in the deeper part of the photoionized gas. The ionizing spectral slope affects the value of $\hnu$ by a factor of $\sim 2$ and thus will not have a significant effect on the above estimates. The metallicity affects the ionization structure, and thus the thickness of the ionized layer. Both effects are explored below in the numerical calculation.

\subsection{Numerical solutions}\label{sec:th_model}
We use the photoionization code Cloudy 10.00 \citep{ferland_etal98} to calculate the structure and line emission of RPC slabs. The code is executed with the `constant pressure' command, which enforces the code to find solutions that satisfy eq.~\ref{eq:dP_dr}. We do not include the contribution of trapped line emission pressure (\Pl) to the gas pressure (see below). The calculation is stopped when the ionized to neutral H fraction drops to 1 per cent. We denote the stopping radius at the back of the slab as $\rb$, and the slab thickness as $d=\rb-\ri$. For some of the models the condition $d/\ri\ll 1$ does not apply, and we therefore always include the geometrical dilution $r^{-2}$ term of the flux within the slab. The total H density (neutral and ionized) at the slab illuminated face \nHi\ is varied in the range $0\leq\log \nHi\leq10$. We explore models in the range $41.5\leq\log L\leq 46$. At $\log L> 45.5$, the condition $d/\ri\leq0.2$ applies in the BLR even for the minimal density explored, and the slab structure is mostly a function of $F_{\rm rad}$, and thus independent of $L$. The largest \ri\ explored is just outside the dust sublimation radius, 
\begin{equation}
 r_{\rm dust}=0.2L_{46}^{0.5}\mbox{~pc}
\end{equation}
\citep{laor_draine93}, i.e.\ twice $r_{\rm BLR}$, where dust suppression of line emission sets the outer boundary of the BLR. The smallest \ri\ explored is $0.03r_{\rm BLR}$, which is close to the size of the optically emitting region in the accretion disc. The specific model explored is with $\log L=45$, where the above range corresponds to $15.5\leq\log \ri\leq 17.5$. The values of metallicity explored are $Z=0.5, 1$ and $5Z_{\sun}$. We adopt the scaling law of the metals with $Z$ from \citet{groves_etal04}. 

Three types of SED are adopted, which differ in the ionizing SED slope \aion\ ($f_\nu\propto\nu^\alpha$). In all cases, the SED is identical between 1~$\mu$m and 1~Ryd. In this range, the SED is evaluated by using
\begin{equation}
f_\nu=\nu^{\alpha_{\rm UV}}\exp(-h\nu/kT_{\rm BB})\exp(-kT_{\rm IR}/h\nu), 
\end{equation}
with $\alpha_{\rm UV}=-0.5$, $kT_{\rm BB}\approx13$~eV and $kT_{\rm IR}\approx0.1$~eV. A cut-off is assumed for $\lambda>1$~$\mu$m. The SED in the 1~Ryd to 1~keV (912--12~\AA) range is fit by a single power-law with $\aion=-1.2$, $-1.6$ and $-2.0$ for the hard, intermediate, and soft SEDs. The adopted \aion\ range corresponds to the observed range of slopes between 1200 and 500~\AA\ \citep{telfer_etal02}. The resulting optical to X-ray slopes are $\aox=-1.16$, $-1.45$ and $-1.74$ for the hard, intermediate and soft SED, respectively, similar to the range observed (e.g.\ \citealt*{brandt_etal00, steffen_etal06}). All three SEDs are extended from 1~keV (12~\AA) up to 100~keV (0.1~\AA) using a single power-law with $\alpha=-1$. A cut-off is assumed above 100~keV. 

Due to some convergence problems in Cloudy, the contribution of \Pl\ to the total $P$ balance is included only in some of the models.\footnote{The description of the convergence problems is available at http://www.nublado.org/ticket/254.} However, we verified that excluding \Pl\ does not change qualitatively the results of our study, as further described in Appendix~\ref{sec:comp_Pline}.

The RPC slab structure typically displays a hot low density surface layer at $\Tcomp $, and a $T\sim 10^4$~K photoionized inner layer, followed by a neutral and colder layer. It must therefore pass through an intermediate photoionized layer at $T\sim 10^5$~K, a layer notorious for its thermal instability (e.g.\ \citealt{krolik99}). The instability allows two separate stable solution with the same thermal pressure, i.e.\ $n_eT$, above and below the unstable part. A similar thermal instability exists in collisionally heated gas (e.g.\ \citealt{draine11b}, ch.~30), where there is a broad range (factor of $>10$) in $T$ of unstable solutions, and thus widely separated combinations of $n_e$ and $T$ of stable solution at a given pressure. In contrast, the ionizing SED of AGN leads to a  more restricted ranges in $T$ of unstable solution, as derived in calculation of X-ray warm absorber models (e.g.\ \citealt{krolik_kriss01, steenbrugge_etal05}). The most complete warm absorber models, which include RPC, are presented by \citet{rozanska_etal06} (see also \citealt{goncalves_etal07, czerny_etal09}). Their model ({\sc titan}) allows to derive the two stable solutions at a given pressure. In contrast, the slab solving scheme  implemented by Cloudy leads to a single solution only. However, at $T<10^5$~K, which is relevant for the BLR, there is only a single stable solution, set by the total pressure in that layer. Cloudy may not be the best tool at the moment to study the structure of the intermediate $T\sim 10^5$~K layer. However, once the cumulative pressure builds up leading to the thermally stable layer at $T< 10^5$~K, the solution should not depend on the exact structure at $T>10^5$~K, a region which is relevant for X-ray warm absorbers.

Below, we adopt models with the intermediate SED (i.e.\ $\aion=-1.6$) and $Z=Z_{\sun}$, unless otherwise noted.

\section{Results}\label{sec:results}

\subsection{The RPC slab structure}\label{sec:results_slab_strc}

\subsubsection{The slab structure versus depth}
\begin{figure*}
\includegraphics[width=174mm]{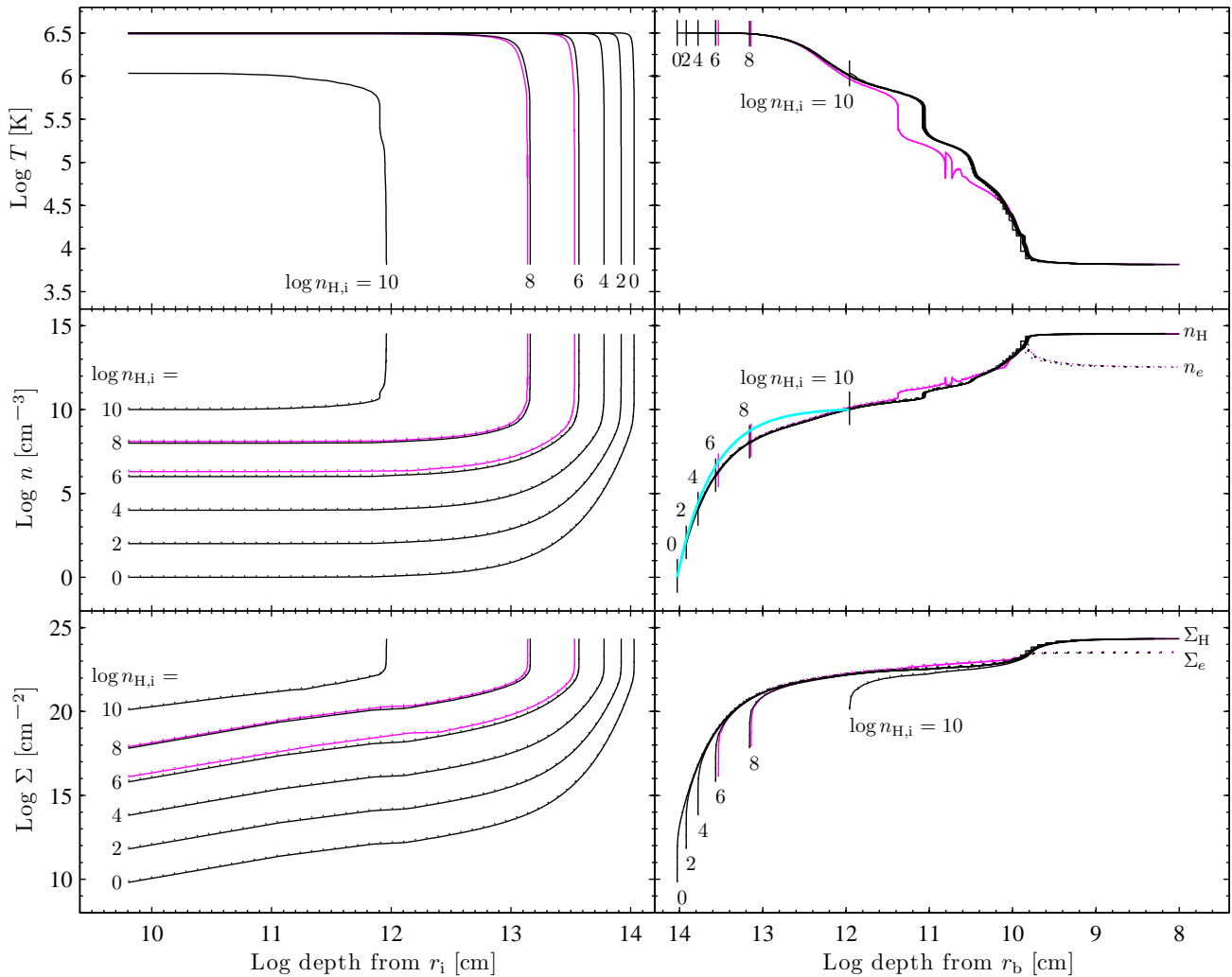}
\caption{The slab structure versus depth for RPC models with $L=10^{45}$~\ergs\ and $\rb=0.03r_{\rm BLR}=10^{15.5}$~cm. Models with different values of $\nHi$ are presented (black lines). Two models which include \Pl\ are also presented for comparison (magenta lines). Left panels present the structure starting from the illuminated face at $\ri$. \emph{Top panel:} The $T$ structure. The RPC slab spans a wide range in $T$, from $\Tcomp=10^{6.5}$~K near $\ri$, down to $T\la10^4$~K near $\rb$. \emph{Middle panel:} The H density and the electron density (solid and dotted line, respectively). A simple analytic model with $n=\nHi\exp{[(r-\ri)/l_{\rm pr}]}$ and $\nHi=1$~\cmt\ (see Appendix~\ref{sec:large_d}) is also presented in the middle-right panel (cyan line). Note that all Cloudy models reach a final $\nH\approx10^{14.5}$~\cmt\ regardless of \nHi. \emph{Bottom panel:} The H and electron $\Sigma$ (solid and dotted line, respectively). Note that the slab remains optically thin to electron scattering ($\tau_{\rm es}\sim0.2$) in all models. The right hand panels present the structure versus distance as measured from $\rb$. Note that on both left and right panels the x-axis starts with the illuminated face on the left side. The tick marks the position of $\ri$ for the different \nHi\ models. The models overlap remarkably well. The slab structure is independent of the value of \nHi, apart from the structure of $\Sigma$ near $\ri$, as $\Sigma$ is an integrated quantity.}\label{fig:struct_nT_15}
\end{figure*}

\begin{figure*}
 \includegraphics[width=174mm]{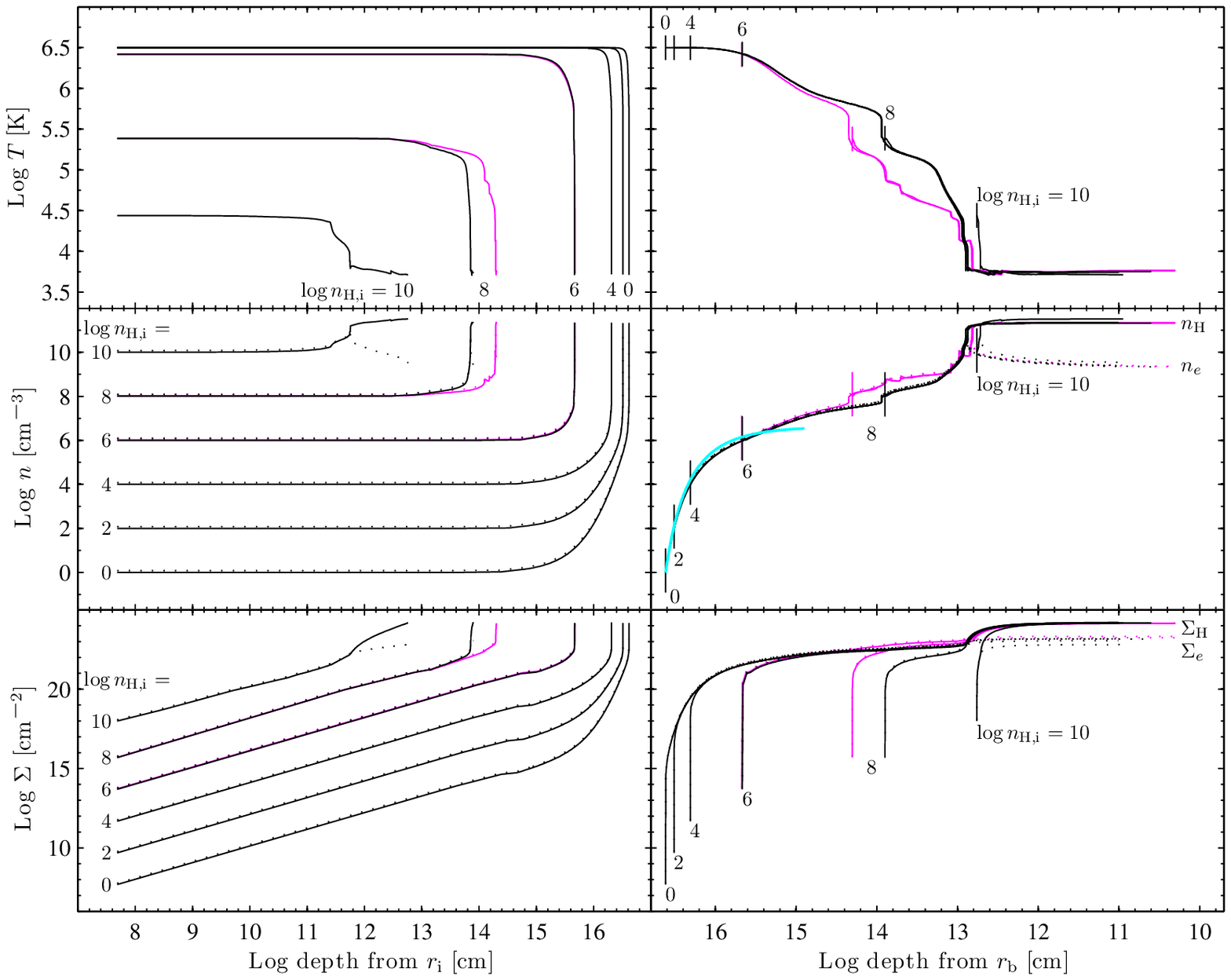}
\caption{The same as Figure~\ref{fig:struct_nT_15}, for a slab at $\rb=r_{\rm BLR}=10^{17}$~cm. The structure is nearly identical to the structure of the slab at $\rb=10^{15.5}$~cm. The x-axis scale is a factor of $10^3$ larger, as expected since $P_{\rm rad}$ is $10^3$ lower. The value of $\nHb$ is $10^3$ smaller, as expected. The range of $T$ and $\Sigma$ is similar to the $\rb=10^{15.5}$~cm model. The analytic model (middle-right panel, cyan line) accounts for the geometrical dilution of flux within the slab (Appendix~\ref{sec:large_d}). Note that the $\log \nH_{,\rm i}=10$ model is not an RPC model since $P_{\rm gas,i}\sim P_{\rm rad}$.}\label{fig:struct_nT_17}
\end{figure*}

Figures~\ref{fig:struct_nT_15} and \ref{fig:struct_nT_17} present $T$, the H density and the electron density $n$, and column density $\Sigma$, in a photoionized RPC slab as a function of the slab depth, for models with $L=10^{45}$~\ergs. Figure~\ref{fig:struct_nT_15}  explores a slab which is situated at $\ri=10^{15.5}$~cm, i.e.\ well inside the BLR (0.03$r_{\rm BLR}$). At this distance $d\ll \ri$, i.e.\ $\ri\simeq \rb$ for all models, and the geometrical dilution is negligible. We explore \nHi\ values in the $1$--$10^{10}$~\cmt\ range. The magenta lines present models that include \Pl, which are described in Appendix~\ref{sec:comp_Pline}. In the left panels the results are presented as a function of distance from $\ri$. The electron temperature $T$ (top left panel) decreases from $\Tcomp\approx10^{6.5}$~K at the illuminated face to $\la10^4$~K at the \Ho\ back side. Note that $d$ is set mostly by the depth of hottest segment of the slab. Once $T$ falls below $\Tcomp$, there is a very sharp drop to $T<10^4$~K due to the sharp rise in the density. The H density (\Ho\ and H$^+$) $n_{\rm H}$ of all models (middle left panel) is nearly constant for an extended region where the gas is fully ionized. As the cumulative optical depth builds up, $P_{\rm gas}$ builds up, leading to a rise in $\nH$. Once $\nH$ is high enough to produce $U<10^3$, the gas is not fully ionized any more, $T$ drops below $T_{\rm C}$, absorption opacity builds up, which leads to a sharper rise in $n_{\rm H}$. The faster rise in $n_{\rm H}$ reduces $U$ more sharply, which leads to further rise in the absorption opacity, and thus to a further increase in the gradient in $n_{\rm H}$. This positive feedback leads to a sharp rise in $n_{\rm H}$ and a sharp drop in $T$. All models reach a density at $\rb$ of $\nHb\approx10^{14.5}$~\cmt, despite the range of $10^{10}$ in \nHi\ of the different models. As expected, since $P_{\rm gas,i}\ll P_{\rm rad}$ for all models, and thus $P_{\rm gas}=P_{\rm rad}$ deep enough where $\tau\gg 1$ and all the radiation is absorbed. The H column density (H$^0$ and H$^+$) $\Sigma_{\rm H}$  rises slowly as long as the cumulative column is low (lower left panel). But, when $\Sigma>10^{22}$~cm$^{-2}$, i.e.\ $\tau_{\rm es}>0.01$, the runaway  rise in $n_{\rm H}$ and the drop in $T$ occurs, and $\Sigma$ also rises sharply. Despite the final high density inside the slab of $\nH\approx10^{14.5}$~\cmt, the slab remains optically thin to electron scattering ($\tau_{\rm es}\approx0.2$) all the way to the inner neutral region, and the use of the Cloudy model (which does not handle $\tau_{\rm es}> 1$) remains valid. The reason that $\tau_{\rm es}> 1$ is not reached is that once $\tau_{\rm es}\sim 0.01$ is reached, $P_{\rm gas}\sim 0.01P_{\rm rad}$, which leads to $U\ll 10^3$. This range of $U$ leads to $\sigma\gg \sigma_{\rm es}$, thus $\tau$ rises rapidly, and $\rb$ is reached within only a small fractional increase in  $\tau_{\rm es}$.

To explore better the structure close to $\rb$, which shows very steep gradients, we plot on the right hand panels all quantities as a function of distance from $\rb$. Remarkably, the different models essentially all overlap. The only difference is the value of  $d$, which increases as \nHi\ decreases. The structure of all models is essentially identical near $\rb$, as expected since $P_{\rm gas}=P_{\rm rad}$, and all models have the same illuminating $P_{\rm rad}$. The value of \nHi\ just sets how far the slab extends inwards from the \Ho\ front, i.e.\ the value of $\rb$. All models have $\nH\approx10^{14.5}$~\cmt\ at $\rb$, since they all have similar $\rb$, i.e.\  similar $P_{\rm rad}$ values. All quantities show a constant value at a distance of $<10^{10}$~cm from $\rb$. This occurs since $\rb$ is defined where the gas is 99 per cent neutral, thus $\tau\gg 1$ and there is no radiative force, and thus no gradient (eq.~\ref{eq:dPdtau}) when getting close to $\rb$. The right-middle panel presents an analytic model with $n=\nHi\exp[(r-\ri)/l_{\rm pr}]$, where $l_{\rm pr}$ is evaluated in Appendix~\ref{sec:large_d}, and $\nHi=1$~\cmt. The analytic model assumes a constant $T=\Tcomp=10^{6.5}$~K, and is therefore truncated at $r$ where $T<10^6$~K. It provides a good match to the detailed photoionization calculations in the hot surface layer.

We thus find that RPC leads to a universal slab structure in the BLR, at a given $P_{\rm rad}$, which is independent of the confining pressure (i.e.\ the value of \nHi). What is the slab structure at a different $P_{\rm rad}$? 

Figure~\ref{fig:struct_nT_17} presents the RPC slab structure for the same models shown in Figure~\ref{fig:struct_nT_15}, but for $\rb=10^{17}$~cm, i.e.\ at $r=r_{\rm BLR}$. In this case, $P_{\rm rad}$ is a factor of $10^3$ lower than in the previous model, which gives $l_{\rm pr}=4\times 10^{15}$~cm (eq.~\ref{eq:r_pr}). Despite the fact that $l_{\rm pr}\ll \rb$ still holds, in the lowest $\nHi\le 10^4$~\cmt\ models, the number of exponential lengths required to reach \nHb\ is $\sim 10$, which leads to $d$ which is only a factor of 2.5 smaller than $\rb$. In the calculation scheme described above, the free parameter is $\ri$, and $\rb$ is a derived quantity. As a result, in the lowest \nHi\ models, similar $\ri$ lead to somewhat different $\rb$, and thus different geometrical dilution effects on $P_{\rm rad}$, which necessarily leads to different \nHb. To avoid this effect, and make a meaningful comparison of the different \nHi\ models, the value of $\ri$ for each model is chosen to yield $\rb=10^{17}$~cm in all models. The graphs have nearly identical shapes to those in Figure~\ref{fig:struct_nT_15}. The main difference is the absolute distance scale, where $d$ is a factor of $10^3$ larger, due to the drop in $P_{\rm rad}$ (eq.~\ref{eq:r_pr}). The values of $T$ and $\Sigma$ are nearly identical, while the \nH\ plot is displaced downward by a factor of $10^3$ ($\nHb=10^{11.3}$ versus $10^{14.5}$~cm$^{-3}$), as it is set by $P_{\rm rad}$. The slight deviation from $\nHb\propto\rb^{-2}$ results from the very high densities reached at $\rb=10^{15.5}$~cm models, and is further discussed below (Section~\ref{sec:final_n}). The only small effect on the shape of the curves in Figure~\ref{fig:struct_nT_17}, compared to Figure~\ref{fig:struct_nT_15}, is the addition of a small correction due to geometric dilution, which leads to a small change in $P_{\rm rad}$ inside the slab. 

The $\log \nH_{\rm ,i}=10$ model (Fig.~\ref{fig:struct_nT_17}) is not a pure RPC model. The model has $P_{\rm gas,i}\simeq0.5P_{\rm rad}$, rather than $P_{\rm gas,i}\ll P_{\rm rad}$, and thus its \nHb\ is larger by $\simeq50$ per cent than for $\log \nH_{\rm ,i}\leq8$ models  where $P_{\rm gas,i}\ll P_{\rm rad}$ (middle panels).

\subsubsection{The slab structure versus $\Sigma_{\rm H}$}
\begin{figure*}
\includegraphics[width=75mm]{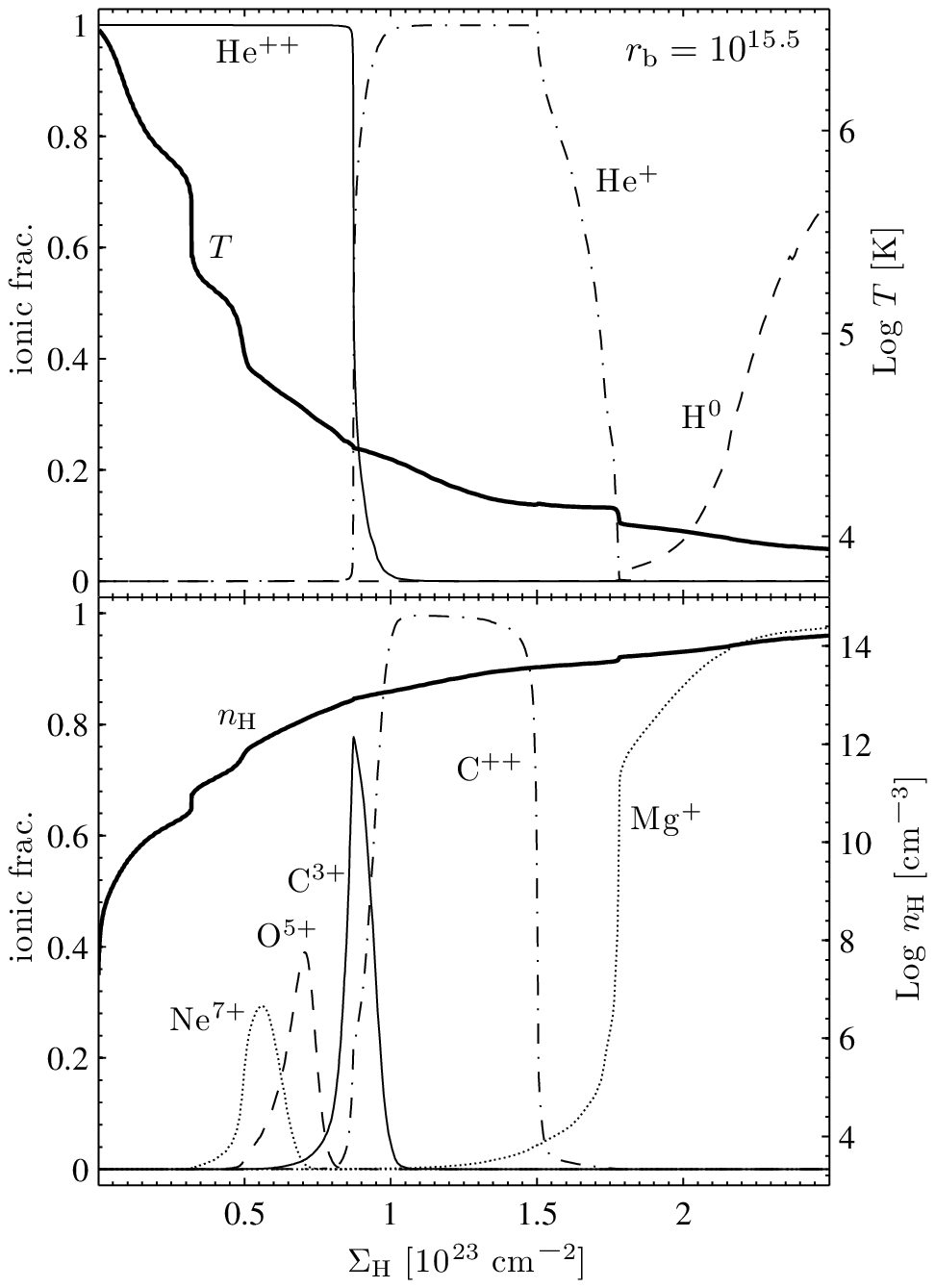}
\includegraphics[width=75mm]{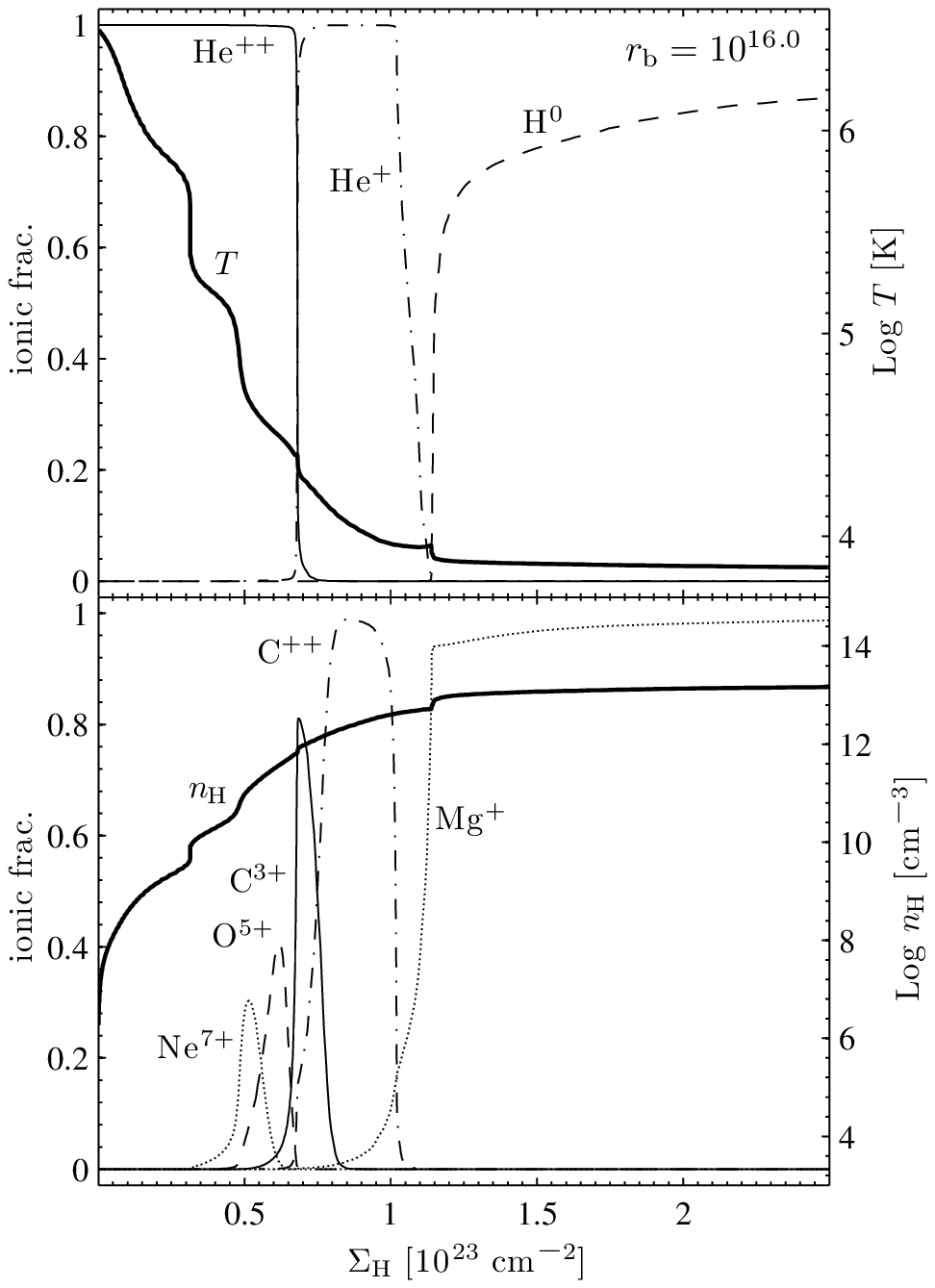}
\includegraphics[width=75mm]{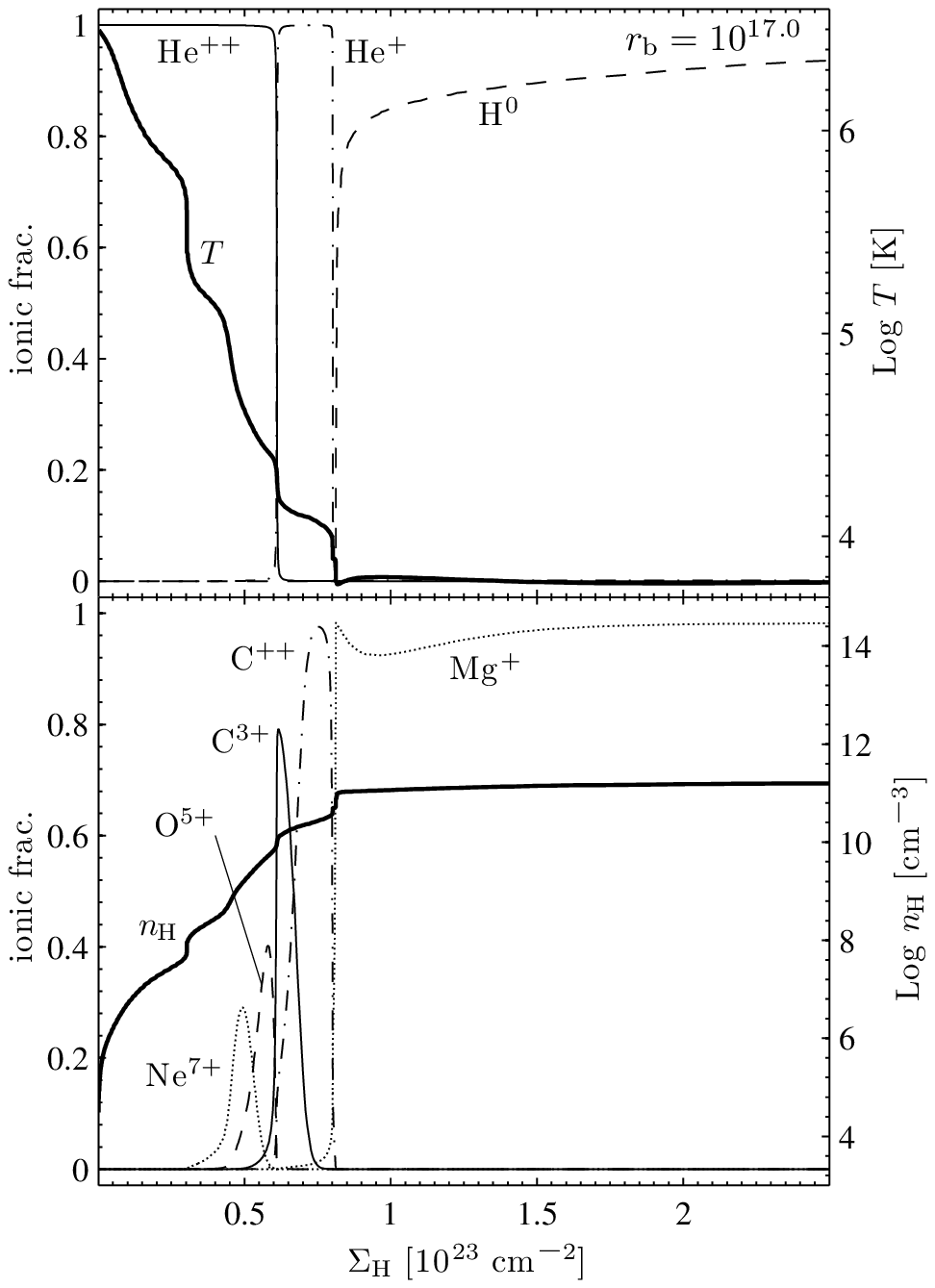}
\includegraphics[width=75mm]{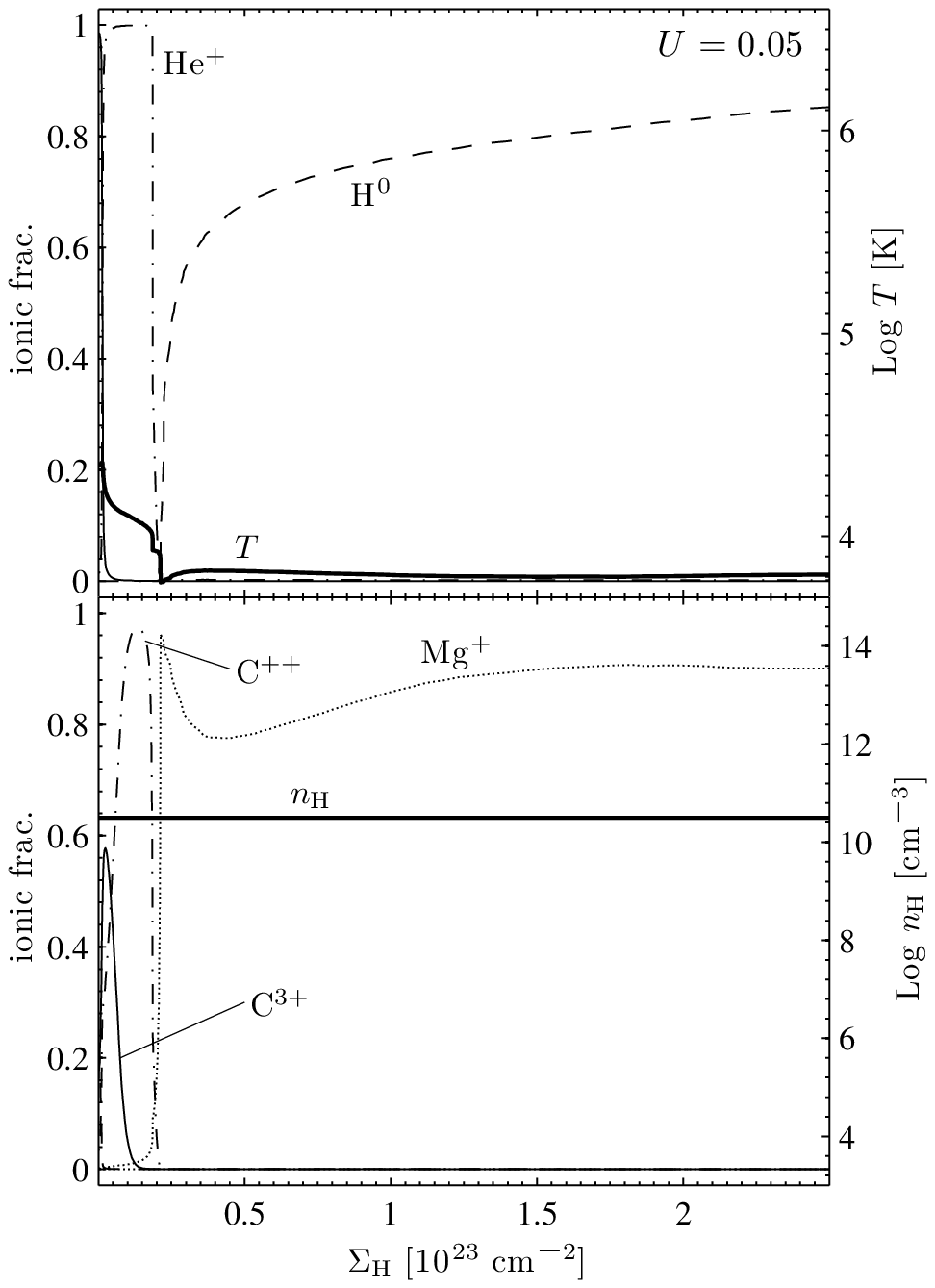}
\caption{The ionization structure of RPC slabs for three $\rb$ values, and a comparison to a uniform density slab. All models assume $L=10^{45}$~erg~s$^{-1}$. Each panel corresponds to a given $\rb$, as noted. The upper left panel corresponds to $0.03r_{\rm BLR}$, upper right to $0.1r_{\rm BLR}$, and the lower left to $r_{\rm BLR}$. The ionization structure is presented as a function of $\Sigma_{\rm H}$. The upper subpanel in each panel presents the $T$ profile (see Figs~\ref{fig:struct_nT_15} and \ref{fig:struct_nT_17}), the fraction of H$^0$, and the ionization fractions of He$^+$ and He$^{++}$. The lower subpanels present the \nH\ profile (see Figs~\ref{fig:struct_nT_15} and \ref{fig:struct_nT_17}), and the ionization fractions of various ions. The lower right panel presents the same plots for the uniform $\nH=10^{10.5}$~\cmt\ model with $U=0.05$. Note the similarity of all RPC models, despite the range of $10^3$ in $F_{\rm rad}$, as expected since RPC sets a universal $U\simeq 0.1$ (see text). The H ionization front and the lower ionization ions (e.g.\ C$^{++}$) extend to higher $\Sigma_{\rm H}$ at smaller $\rb$, while the higher ionization ions (e.g.\ Ne$^{7+}$ and O$^{5+}$) occur at similar $\Sigma_{\rm H}$ at all $\rb$. The uniform \nH\ model (lower right panel) shows only a small range in $T$, lacks Ne$^{7+}$ and shows only little O$^{5+}$. The RPC slab can be viewed as a superposition of matter bounded fixed-$U$ slabs for $U$ in the range of 1--1000, and an ionization bounded $U\sim 0.1$ slab.}\label{fig:struct_all}
\end{figure*}

Figure~\ref{fig:struct_all} shows the ionization structure of RPC slabs, as a function of $\Sigma_{\rm H}$, at three distances $\rb=r_{\rm BLR}$, $\rb=0.1r_{\rm BLR}$ and $\rb=0.03r_{\rm BLR}$, which correspond to $10^{17}$, $10^{16}$ and $10^{15.5}$~cm for $L=10^{45}$~erg~s$^{-1}$. The figure also provides a comparison of RPC to a uniform density slab at $U=0.05$. All RPC solutions assume $\nHi=10^2$~\cmt, but as noted above the solution is not sensitive to this value. Each panel presents the results for a certain $\rb$, and each panel is divided into two subpanels. The top subpanel presents $T$ and the ionization fractions of H$^0$, He$^+$ and He$^{++}$. The lower subpanel  presents \nH\ and the ionization fraction of several high ionizations ions (Ne$^{7+}$ and O$^{5+}$), intermediate ionizations (C$^{3+}$ and C$^{++}$) and low ionizations (Mg$^+$). The structure is similar for the three RPC models, despite the range of $10^3$ in $F_{\rm rad}$. The main trend observed is the decrease in $\Sigma_{\rm H}$ of the H ionization front (say where the H ionization is 50 per cent) with increasing $\rb$. The position drops from $\Sigma_{\rm H}=2.3\times 10^{23}$~\cmmt\ at $\log \rb=15.5$, to $0.8\times 10^{23}$~\cmmt\ at $\log \rb=17$. This also roughly corresponds to the trend in the column where $T$ reaches $10^4$~K. However, the column were $T$ drops to $10^5$~K remains close to $0.5\times 10^{23}$~\cmmt\ for all $\rb$ values. Similarly, the Ne$^{7+}$ ion peaks at $\Sigma_{\rm H}=0.5\times 10^{23}$~\cmmt\ at all $\rb$, while the mean position of the C$^{++}$ ion drops from $1.2\times 10^{23}$ to $0.75\times 10^{23}$~\cmmt, with the increase in $\rb$. The Mg$^+$ ion rises sharply at the H ionization front for all \rb, and remains nearly constant and close to unity beyond the H ionization front. The models do not reach the $\Sigma_{\rm H}$ where Mg$^0$ becomes dominant,  as this ion is ionized by photons above 0.6~Ryd, where $\tau\ll1$ for $\Sigma_{\rm H}\leq 10^{24}$~\cmmt. The shift of the H ionization front with $\rb$ is a density effect, and not an RPC effect. It occurs also for uniform density models, where models with a higher $n$ value, at a fixed $U$, yield larger $\Sigma_{\rm H}$ values to reach the H ionization front.

Figure~\ref{fig:struct_all}, lower right panel, shows a comparison of RPC to a uniform density slab model with $U=0.05$, which corresponds to $\nH=10^{10.5}$~\cmt, expected at $\rb=10^{17}$~cm. This \nH\ value is where C$^{3+}$ and C$^{++}$ ionic fractions peak in the RPC solution for $\rb=10^{17}$~cm (lower left panel). The major difference, apart from the lack of a density gradient, is the $T(\NH)$ profile, which now extends only over a small range in $T$ near $10^4$~K. The H ionization front now occurs at a column of $0.3\times 10^{23}$~\cmmt, rather than $0.8\times 10^{23}$~\cmmt. This difference is expected since the ionized column is proportional to $U$, and in an RPC slab $U$ increases from $\rb$ to $\ri$, while in a uniform density slab $U$ is constant. The difference associated with the lack of rise in $U$ towards $\ri$, is the lack of a region with Ne$^{7+}$ emission, and the fact that O$^{5+}$ is present only at a thin surface layer. As can be seen in the other panels with the RPC solutions, the O$^{5+}$ ion peaks where \nH$\sim 0.1$\nHb, i.e.\ where $U\sim 1$, while Ne$^{7+}$ peaks where \nH$\sim 0.01$\nHb, i.e.\ at $U\sim 10$. Because of the density gradient within an RPC slab, it can be viewed effectively as a superposition of matter bounded (i.e.\ optically thin), constant $U$ slabs, with $U$ in the range of 1--1000, and in addition an ionization bounded $U\simeq 0.1$  slab. Although a constant $U$ slab also displays an ionization gradient within the slab, as the ionizing photons absorption increases inwards, an RPC slab shows a larger range of $U$ values, because of the gradient in \nH.

\subsubsection{The slab final density versus \rb\ and $L$}\label{sec:final_n}
\begin{figure}
 \includegraphics[width=84mm]{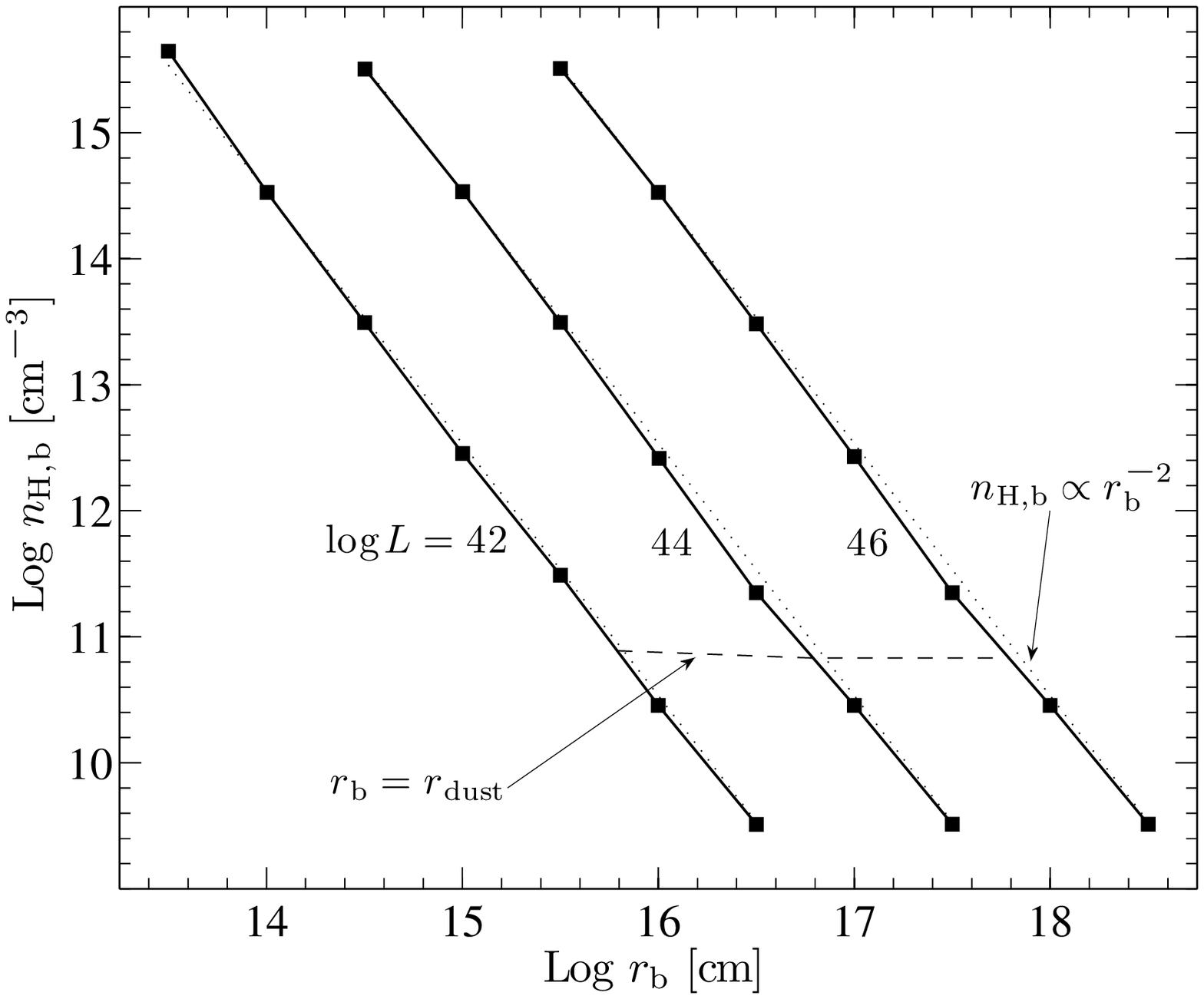}
\caption{The density at the inner part of an RPC slab versus \rb. The photoionization model results are presented for $L=10^{42}$, $10^{44}$ and $10^{46}$~\ergs\ (solid line with square markers; each square marker denotes a specific photoionization model run). Each \nHb\ value corresponds to a specific $F_{\rm rad}$, and the value which corresponds to dust sublimation is indicated (dashed line). This flux corresponds to a distance $\rb=r_{\rm dust}=0.2L_{46}^{1/2}$~pc \citep{laor_draine93}. An analytic fit is also presented (dotted lines). Note the overall good agreement between the analytic fit and photoionization models. At $\nHb<10^{10.8}$~\cmt\ dust can survive, which leads to a jump of $\sim1.5$ in $P_{\rm rad}$, as non ionizing continuum is also absorbed by dust, and therefore contributes to $P_{\rm rad}$ (see text).}\label{fig:nf}
\end{figure}

Figure~\ref{fig:nf} presents the derived \nHb\ from Cloudy model solutions as a function of \rb\ for $L=10^{42}$, $10^{44}$ and $10^{46}$~\ergs. The value of \nHb\ rises linearly with $L$ at a given \rb, and drops as $\rb^{-2}$ at a given $L$. Thus, $\nHb\propto L/\rb^2$, i.e.\ $\propto P_{\rm rad}$, as expected. The importance of RPC is that it gives an absolute value for \nHb\ for given $L$ and \rb\ values, with no free parameters. The curves shown in Fig.~\ref{fig:nf} are well fit by the universal relation 
\begin{equation}
\nHb=3\times 10^{14}L_{46}\,r_{16}^{-2},
\end{equation}
where $r=10^{16}r_{16}$~cm. Since $r_{\rm BLR}\approx 0.1L^{0.5}_{46}$~pc, we get the simple relation
\begin{equation}
\nHb\approx 2.5\times 10^{11}(r/r_{\rm BLR})^{-2},
\end{equation}
which fits well the Cloudy derived \nHb, as shown in Fig.~\ref{fig:nf}. This fit can also be derived from the simple analytic model of \nHb, where
\begin{equation}
 \nHb=L/4\pi\rb^2ckT,
\end{equation}
using $T=10^{3.75}$~K which corresponds to the slab neutral face (e.g.\ Fig.~\ref{fig:struct_all}). The slight deviation from $\nHb\propto\rb^{-2}$ present in Cloudy models for large \rb, which approaches $r_{\rm dust}$ from below, can be explained as follows. At small \rb, \nH\ is very large ($\geq10^{13}$~\cmt), and the gas is efficient in absorbing  at $\lambda>912$~\AA, due to photoionization of excited H. At larger \rb, \nH\ is smaller, and the gas is inefficient in absorbing longward of 912~\AA. This effectively decreases the absorbed fraction of $L$, i.e.\ decreases the effective $P_{\rm rad}$ which sets \nHb. At $\rb\geq r_{\rm dust}$, dust absorption sets in, the nonionizing continuum ($\lambda>912$~\AA) contributes again to $P_{\rm rad}$, which leads to a factor of $\sim1.5$ jump in \nHb\ (see also Paper I).

\subsection{The predicted line emission}\label{sec:results_line_em}
\begin{figure*}
\includegraphics[width=150mm]{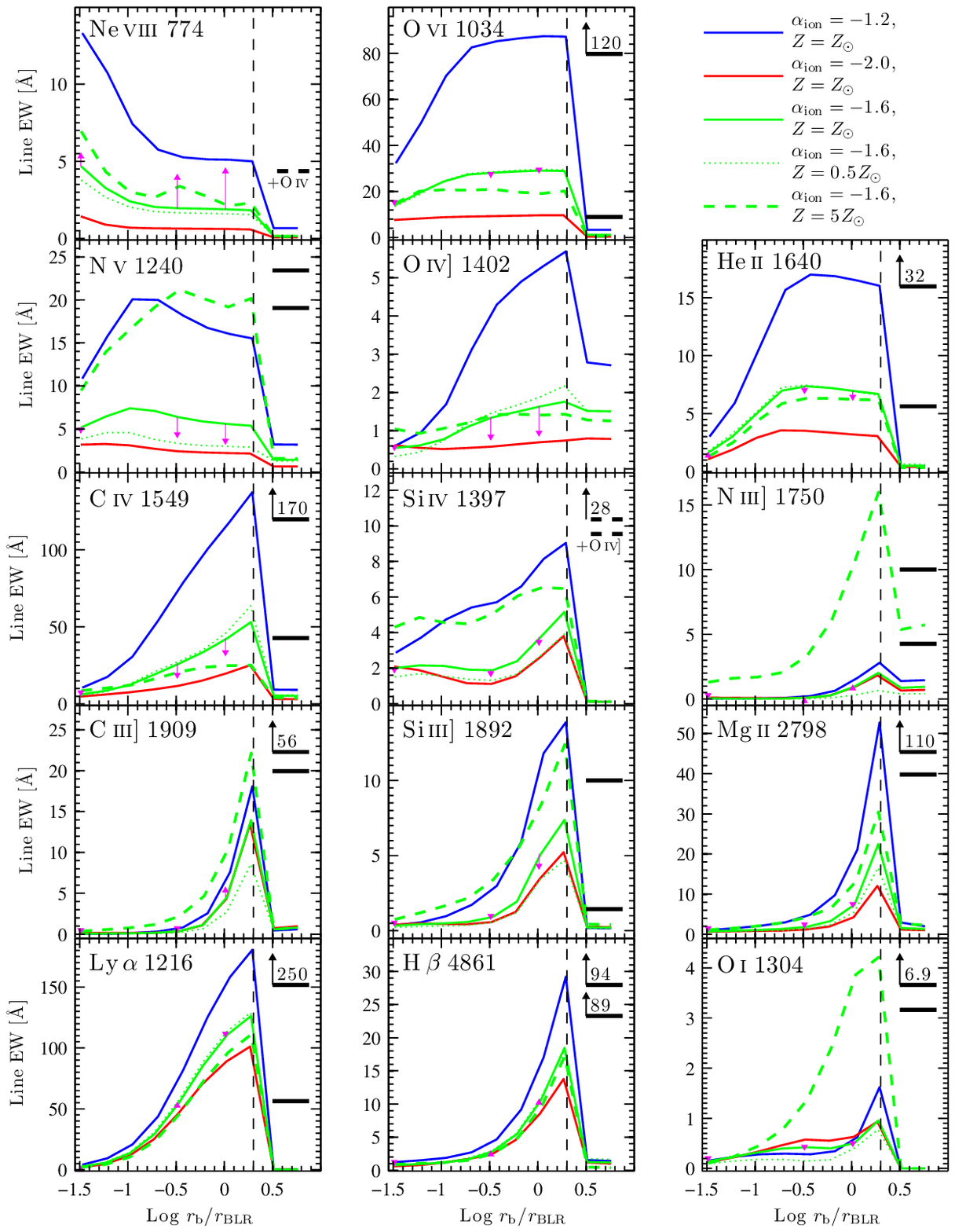}
\caption{Calculated line EW versus $\rb/r_{\rm BLR}$ for RPC models with $L=10^{45}$~\ergs\ and $\Omega_{\rm BLR}=0.3$. The line EW is evaluated for a range of values of $\alpha_{\rm ion}$ and $Z$, noted in the upper-right corner. The dust sublimation radius $r_{\rm dust}$ is indicated (vertical dashed line). Models with $\rb>r_{\rm dust}$ include the effect of dust opacity, which strongly suppresses the line emission. The effect of including \Pl\ in models with $\alpha_{\rm ion}=-1.6$ and $Z=Z_{\sun}$ is depicted by a vertical arrow at $\log\rb/r_{\rm BLR}=-1.5$, $-0.5$ and 0. The thick horizontal bars (right hand sides) depict the measured line EW in objects with $\lambda L_{\lambda}(1450\mbox{\AA})=10^{42}$ and $10^{47}$~\ergs\ from \citet{dietrich_etal02} (\NeVIII\ EW is from \citealt{telfer_etal02}). The measured EW of emission blends is depicted by a dashed bar, and the additional contributing emission line is denoted below the lowest bar. Bars with an arrow denote measured EW values which are outside the plotted range, and the measured value is noted next to the arrow. In the RPC photoionization models, the low ionization lines mostly originate within a factor of 2--3 of $r_{\rm dust}$, i.e.\ at $\rb\simeq r_{\rm BLR}$. At intermediate ionizations, the line emission extends to a smaller $\rb$, and the highest ionization lines either peak at $\rb\simeq 0.1r_{\rm BLR}$, or at \rb\ smaller than $0.03r_{\rm BLR}$. If the inner radius of the BLR is set by a failed dusty wind (\citealt{czerny_hryniewicz11}), it may not extend to such a small $\rb$.}\label{fig:EW}
\end{figure*}

Figure~\ref{fig:EW} presents the calculated emission-line equivalent-width (EW) versus $\rb/r_{\rm BLR}$ for several prominent lines from the BLR. A BLR covering factor of $\Omega_{\rm BLR}=0.3$ is assumed. The results are independent of the value of $\nHi$ (assumed $10^6$~\cmt\ here). The EW is calculated adopting several values of \aion\ and $Z$ (Section~\ref{sec:th_model}). Models with $\rb>r_{\rm dust}$ include the effect of dust opacity and metal depletion on grains (assuming an ISM-like depletion). Since RPC sets the structure of $U$ and \nH\ in the photoionized layer at a given $\rb$, the predicted EW is unique. The only free parameters are  $\dd\Omega_{\rm BLR}/\dd r$, \aion\ and $Z$. For a `typical' SED with \aion$=-1.6$ and $Z=Z_{\sun}$ we find following. The emission of the low ionization lines, in particular H$\beta$, \MgII~$\lambda$2798 and \CIII]~$\lambda$1909 is sharply peaked at $r_{\rm dust}$. The sharp drop at $\rb>r_{\rm dust}$ is a universal effect which affects most lines, as expected since dust dominates the opacity for $U> 0.01$ \citep{netzer_laor93}, while RPC produces $U\ge 0.1$. The EW of these lines drops by a factor of 3--10 when \rb\ becomes smaller than $r_{\rm dust}$ by a factor of 2--3. Thus, the low ionization lines are mostly confined to a narrow range of radii close to $r_{\rm BLR}$ ($=0.5r_{\rm dust}$). Intermediate ionization lines, such as \CIV~$\lambda$1549, \OIV]~$\lambda$1402 and \SiIV~$\lambda$1397, still peak at $r_{\rm dust}$, but the EW drop occurs at a smaller $\rb$. Higher ionization lines, such as \HeII~$\lambda$1640 and \OVI~$\lambda$1034, maintain a high EW down to $\rb\sim 0.1r_{\rm BLR}$, the \NV~$\lambda$1240 line peaks at $\sim 0.1r_{\rm BLR}$, while the EW of the \NeVIII~$\lambda$774 line keeps increasing to the limit of the calculation at $0.03 r_{\rm BLR}$. The physical mechanism that produces the dependence of EW on $r$ is discussed in Section~\ref{sec:disc_rad_dist}.

A harder ionizing SED (i.e.\ $\aion=-1.2$) increases the EW of all lines, as expected since more energy is absorbed and re-emitted in lines, for a given near UV luminosity (used to calculate the EW of most of the UV lines). The effect is particularly strong for the higher ionization lines, as expected since a harder EUV SED produces a larger column of high ionization ions. The effect of $Z$ is more complicated. The line emission enhancement is particularly strong for the N lines, such as the \NIII~$\lambda$1750 and \NV~$\lambda$1240 lines presented here, as expected given the steeper scaling of N with $Z$ (Section~\ref{sec:th_model}). The H lines, and the \HeII~$\lambda$1640 are almost independent of $Z$, as these lines mostly count the number of ionizing photons. Prominent coolants, such as \CIV~$\lambda$1549 and \OVI~$\lambda$1034 get weaker with increasing $Z$, which happens as these lines are collisionally excited, and photoionized gas gets colder with increasing $Z$ due to increased cooling by the many weaker lines. Note for example the sharp increase in the \OI~$\lambda$1304 line with $Z$. This rise characterizes other weak low ionization lines not plotted (e.g.\ \CII~$\lambda$1335).

\subsubsection{Comparison with observations}
Figure~\ref{fig:EW} also presents a comparison with the typical observed EW of the various lines. Since most lines present a Baldwin effect, the typical EW is a function of luminosity. We used the best fit EW versus luminosity relation in \citet{dietrich_etal02}, and derived the EW values at $\lambda L_{\lambda}(1450\mbox{\AA})=10^{42}$ and $10^{47}$~\ergs\ where the lower values of the measured EW correspond to  high $\lambda L_{\lambda}$ (except for \NV).\footnote{The EW of \NeVIII\ is not reported in \citet{dietrich_etal02}. We present in Fig.~\ref{fig:EW} the EW value of \NeVIII+\OIV\ blend that was measured by \citet{telfer_etal02} for a composite spectrum with a mean $\lambda L_{\lambda}(1100\mbox{\AA})=10^{46}$~\ergs. The \OIV~$\lambda$789 contribution to the blend EW is likely to be small, and it is $<10$~per cent of the \NeVIII\ EW in the RPC models.} The calculated and measured  EW are generally consistent, for the adopted $\Omega_{\rm BLR}=0.3$, as derived in earlier studies \citep{korista_etal97, maiolino_elal01, ruff_etal12}. 

There is a clear trend of EW with $\aion$ from the RPC calculations, as found in earlier photoionization modelling of uniform density gas \citep{korista_etal98}. The observed trend of the EW with luminosity likely results from the trend of a softer SED with increasing luminosity (e.g.\ \citealt{steffen_etal06}), as suggested by \citet{korista_etal98}.

The predicted emission of \NIII\ and \OI\ is a strong function of $Z$ only, and a comparison with observations implies $Z>Z_{\sun}$ for the BLR, which is consistent with previous studies \citep{hamann_ferland93, hamann97, hamann_ferland99}. The \NV\ line either implies $Z>Z_{\sun}$ for a `typical' SED ($\aion\sim-1.6$), or $Z\sim Z_{\sun}$ for a hard SED ($\aion\sim-1.2$). This degeneracy might explain the lack of a Baldwin relation for \NV, as further discussed below (Section~\ref{sec:disc_obsr_trends}).

The emission of several lines (\CIII], \MgII) is significantly underpredicted (factor of $>2$) by RPC models for low-$L$ objects (i.e.\ high EW and shallow \aion). The \CIII] line is measured by decomposing the \CIII]+\SiIII]+\AlIII\ emission blend. To minimize the effect of decomposition uncertainties, we make a comparison between the total calculated and measured EW of the blend. The measured EW is $\simeq80$~\AA\ \citep{dietrich_etal02}, which is still significantly higher compared to 34~\AA\ predicted by RPC models for $\aion=-1.2$ (\AlIII\ contributes $\simeq2$~\AA). The discrepancy between the two values might be explained, if the drop in line emission beyond $r_{\rm dust}$ is more gradual than in our models (Section~\ref{sec:disc_rad_dist}). The gas at $\rb\ga r_{\rm dust}$ is prone to emit \CIII] more efficiently than at $\rb< r_{\rm dust}$, if the emission is not suppressed by dust, since the gas has $\nHb\la10^{10.5}$~\cmt\ (Fig.~\ref{fig:nf}), which is closer to the critical density of \CIII] ($10^{9.5}$~\cmt). A gradual drop in line emission beyond $r_{\rm dust}$ might also explain the discrepancy between the calculated and measured EW of \MgII. 

The high value of the \OVI\ EW (120~\AA) is an extrapolation to $\lambda L_\lambda=10^{42}$~\ergs\ of a relation evaluated by \citet{dietrich_etal02} only down to $\lambda L_\lambda\simeq10^{44}$~\ergs. Since the slope of \OVI\ EW versus $L$ relation is likely to be shallower for $L\la10^{44}$~\ergs, as it is for other high ionization lines (e.g.\ \CIV; \citealt{dietrich_etal02}), the \OVI\ EW at $L\sim10^{42}$~\ergs\ is probably closer to $\la70$~\AA\ (fig.~7 there), and is consistent with RPC models. 

The \SiIV\ panel in Fig.~\ref{fig:EW} presents the measured total EW of \SiIV+\OIV] blend. The total calculated EW of \SiIV+\OIV] is roughly consistent with these observed values.

The calculated \Hb\ EW is significantly smaller for all models than observed (by a factor of $\ga3$). This discrepancy is likely related to the famous and persistent `\Lya/\Hb\ problem' \citep{baldwin77a, netzer_etal95}, which may result from an underestimate of the \Hb\ emission in photoionization calculations, potentially due to a radiative transfer effect (e.g.\ \citealt{baldwin_etal04}, section~8.4 there for a possibly related effect in the underestimate of the optical \FeII\ emission).

\section{Discussion}\label{sec:discuss}

As shown in earlier studies \citep{laor_draine93, netzer_laor93, korista_etal97} and verified by observations \citep{suganuma_etal06}, the BLR is bounded from the outside by dusty photoionized gas, as dust embedded in the line emitting gas strongly suppresses the line emission once $U>10^{-2}$. Dust sublimation explains both the scaling and absolute value of the observed $r_{\rm BLR}\approx 0.1L^{0.5}_{46}$~pc relation. Dust also provides a natural inner boundary for the BLR, if the BLR gas originates from a failed dusty wind, and the wind exists down to the sublimation radius within the accretion disc atmosphere, which is smaller by a factor of few of the dust sublimation radius above the disc \citep{czerny_hryniewicz11}. These mechanisms imply that the BLR gas is subject to a characteristic $F_{\rm rad}\simeq 10^{10}$~erg~s$^{-1}$~cm$^{-2}$, which corresponds to $P_{\rm rad}\simeq 0.3$~erg~cm$^{-3}$, or an ionizing photon density $n_{\gamma}\simeq6\times 10^9$~cm$^{-3}$. 

In this study we point out that $P_{\rm rad}$ inevitably leads to a pressure gradient within the photoionized gas. The thermal pressure within the photoionized gas therefore builds up to $P_{\rm gas}=P_{\rm rad}$, or $2n_ekT=n_{\gamma}\hnu$, which implies $n_e\simeq 15 n_{\gamma}$. Thus, RPC leads to a fixed ratio of $U\simeq 0.1$ for the typical AGN SED, as found by  observation. These two mechanisms, dust which sets $n_{\gamma}$, and RPC which sets $U$, lead to a unique solution for the BLR structure and line emission, independent of $L$. These mechanisms provide a natural explanation for the similar BLR emission observed over a range of $10^8$ in AGN luminosity. 

An RPC slab of gas is characterized by a density which increases inwards, reaching \nHb\ once all the ionizing $F_{\rm rad}$ is absorbed (which can include radiation longwards of 912~\AA,  depending on the ionization and excitation states). This generally implies $\nHb\propto r^{-2}$ in RPC gas near a point source.  The density structure of the photoionized layer, as measured from the neutral inner part, has a nearly universal shape. The boundary gas pressure $P_{\rm gas,i}$ at the surface of the ionized gas has no effect on the inner structure, as long as  $P_{\rm gas,i}\ll P_{\rm rad}$. The structure is characterized by a surface layer at $\Tcomp \sim 10^{6.5}$~K for the typical AGN SED, which exists in regions where $\nH<10^{-6}\nHb$ (Figs~\ref{fig:struct_nT_15} and \ref{fig:struct_nT_17}). This hot surface layer exists at $\tau_{\rm es}<10^{-2}$, where $P_{\rm gas}<10^{-2}P_{\rm rad}$, which implies $U>10^3$, required for a fully ionized gas. Once $\tau_{\rm es}>10^{-2}$ inside the cloud, one gets $U<10^3$, bound electrons become abundant, the absorption opacity rises, leading to a warm, $T\sim 10^5$~K, and possibly thermally unstable layer. This layer may break up into two phases according to the stable regions in the $P_{\rm gas}$ versus $T$ photoionization solutions, which we do not explore in this study. However, once $\tau_{\rm es}\ga 0.1$, one gets $T\sim 10^4$~K, the bulk of the ionizing luminosity is absorbed, and the prominent optical and UV BLR lines are produced.

The ionization structure and the line emission from RPC gas is equivalent to a superposition of optically thin uniform density slabs, with $U$ decreasing from 100 (or higher) to 1, and an additional ionization bounded slab with $U\sim 0.1$. Thus, the main difference between the standard BLR solution of a uniform density slab, and the RPC solution, is the addition of higher ionization `surface' layers. These layers are responsible for excess \OVI~$\lambda$1034 and \NeVIII~$\lambda$774 emission, which is otherwise interpreted as additional BLR cloud components with $U\sim 1$ and $U\sim 10$ \citep{netzer76, korista_etal97, hamann_etal98}. RPC predicts there should also be an additional BLR component of even higher ionization lines, produced by the $U\sim 100$ layer. This lower density layer may produce some forbidden lines from highly ionized ions, however the emission may be hard to detect, as this layer absorbs only $\sim 1$ per cent of the ionizing radiation, and the predicted line luminosities are correspondingly weak. In addition, the main coolants occur in the extreme UV, such as the various Fe lines: \FeXXIII~$\lambda$133, \FeXXII~$\lambda$111 and \FeXX~$\lambda$121, and are therefore essentially unobservable. The associated forbidden optical and NUV lines from these ions are expected to produce very weak broad lines, and are thus very difficult to detect. 

\subsection{The stability of RPC BLR}\label{sec:disc_stability}

Is the hydrostatic solution maintained by radiation pressure stable? Earlier studies of radiation pressure effects on photoionized gas \citep{mathews_blumenthal77, mathews82, mathews86} found that radiatively driven photoionized gas may be susceptible to various instabilities. These studies were motivated by early suggestions to explain the broad line profiles as a radiatively driven outflow. However, we now know that the BLR dynamics is largely dominated by the gravity of the black hole, which indicates it must be composed of a large column gas, where gravity inevitably dominates radiation pressure. We also know that the BLR gas is not composed of small clouds \citep{arav_etal98, laor_etal06}, and is most likely a smooth flow, possibly formed by a failed disc wind \citep{czerny_hryniewicz11}. Thus, the earlier studies are not relevant for the BLR we know today. We do not attempt to explore the stability of the hydrostatic solution. We can only comment that the photoionized layer must be supported by a large column ($>10^{24}$~cm$^{-2}$) of static gas on the back side, to avoid forming a wind, and we suspect this support can allow a stable solution. 

\subsection{The validity of a steady state solution}\label{}

Is the steady state solution a valid assumption? We know the ionizing continuum in AGN varies with time. How long will it take for the RPC solution to adapt to a new $L$? Even if $L$ is a constant, the BLR gas likely rises from the accretion disc, enters the ionizing radiation field, and needs to reach a steady state structure. The physical thickness of the `proper' BLR layer, i.e.\ where $n>10^9$~\cmt, is $<10^{13}$~cm (Fig.~\ref{fig:struct_nT_17}). If the rising gas is compressed by the radiation pressure, the minimal radial acceleration is  $F_{\rm rad}\sigma_{\rm es}/cm_{\rm p}\sim 0.1$~cm~s$^{-2}$, which gives a maximal time of $\sqrt{10^{13}/0.1}=10^7$~s. However, since the flux weighted mean absorption opacity in BLR gas is likely $10^3\sigma_{\rm es}$, the compression timescale can be a few days, rather than a few months. If the gas rising from the accretion disc has internal pressure larger than implied by RPC, it will expand at the sound speed of $c_{\rm s}\sim 10^6$~cm~s$^{-1}$ and will fill in the required width in $10^{13}/10^{6}=10^{7}$~s. The dynamical timescale (\tdyn) at the BLR is $\sim 100\times$ the light crossing time (assuming a Keplerian velocity of $0.01c$), and thus typically at least few years. A similar timescale is obtained if the vertical motion from the disc to the BLR is a fraction of the Keplerian velocity, and the gas rises above the disc to a fraction of the radius. We therefore expect that the travel time of the gas from the disc surface to the height of the BLR, is comparable to the time it will take the gas to expand and relax to the steady state RPC solution. 

The above considerations also apply for luminosity variability. When the luminosity increases, the BLR density will adapt to the higher density RPC solution on a short timescale, likely a few days. However, when the luminosity decreases, the relaxation to the lower density solution occurs only at the sound crossing time, i.e.\ $10^7$~s. One may therefore expect some hysteresis effect in the response of the line emission to continuum variations.

\subsection{The effect of  $\Sigma_{\rm H, total}< 10^{24}$~\cmmt}\label{sec:disc_smaller_nh}

If the total H column of the gas ($\Sigma_{\rm H, total}$) is $< 10^{24}$~\cmmt, then our assumption that gravity dominates the radiative force may not hold. Specifically, the ratio of the radiation force to gravity for an optically thick cloud is $R\equiv 1.5l\Sigma_{24}^{-1}$, where $l=L/L_{\rm Edd}$ is the luminosity in Eddington units, and $\Sigma_{24}=\Sigma_{\rm H, total}/10^{24}$~\cmmt. If $R>1$, e.g.\ for gas with $\Sigma_{24}<0.1$ in moderately high, $l>0.1$, AGN, the gas will be radially accelerated and form an outflow. The outflowing gas may still be confined by RPC, due to the differential acceleration of different layers within the gas cloud, relative to the bulk acceleration of the cloud (e.g.\ \citealt{chelouche_netzer01}). However, since the BLR is generally not an outflowing wind, gas clouds with $R>1$ do not form the bulk of the BLR emitting gas. 

If $R\ll 1$, then the hydrostatic solution derived above is valid. The gas slab structure is presented in Figure~\ref{fig:struct_all} up to a column of $\NH=0.25\times10^{24}$~\cmmt, but the solution extends to $\Sigma_{24}=1$, where the neutral H fraction equals 1 per cent (Sec.~\ref{sec:th_model}). The calculations are valid for AGN with $l<1$, if $\Sigma_{24}>1$.  A hydrostatic solution for a $\Sigma_{24}=0.1$ cloud remains valid for AGN with $l\ll 0.1$. The effect of reducing $\Sigma_{\rm H, total}$ to $10^{23}$~\cmmt\ on the predicted line emission is small. We recalculate the line emission from an RPC slab located at $\rb=r_{\rm BLR}$, and now stop the calculation when  $\Sigma_{24}=0.1$, rather than at $\Sigma_{24}=1$. The emission of \Hb\ decreases by $\sim20$ per cent, and that of \NIII]~$\lambda$1750 and \OI~$\lambda$1304 increases by $\sim10$ and $60$ per cent, respectively. The emission of all other lines, presented in Fig.~\ref{fig:EW}, differs by less than 1 per cent. Note that an RPC slab achieves its full ionization stratification at $\NH\la 2\times10^{23}$~\cmmt\ (Fig.~\ref{fig:struct_all}), and thus BLR clouds with $\Sigma_{\rm H, total}$ of a few $10^{23}$~\cmmt\ \citep{risaliti_etal11} are likely to contain emitting ions of all ionization states.

What happens at the intermediate case when $R\la 1$? In this case, the gas can still maintain a circular orbit, but the radial acceleration will be reduced significantly by the radiation force, which opposes a significant fraction of the black hole gravity \citep{marconi_etal08, marconi_etal09}. A hydrostatic solution is still valid in the rotating frame. However, the gas pressure gradient now balances only the differential radiative acceleration between the bulk acceleration of the cloud and the local radiative acceleration. For example, if $R=0.9$, then the gas pressure will build up to a maximal value of $0.1$ of the value for the $R\ll 1$ solution, which corresponds to $U\sim 1$, rather than $U\sim 0.1$. However, one needs fine tuning of $R$ to be just below unity to have a significant effect, and such clouds will likely be rare.

\subsection{The effect of the shear}\label{sec:disc_thick}
Above (Section~\ref{sec:theory}) we considered the geometrical dilution of the incident flux in a geometrically thick slab, i.e.\ a slab where $d\equiv\rb-\ri$ is comparable to \rb. Another significant effect is a shear due to the gradient in the Keplerian velocity with $r$. The difference in Keplerian velocity across the slab thickness is 
\begin{equation}
\Delta v_{\rm K}\simeq \frac{1}{2}\frac{\sqrt{G M_{\rm BH}}}{r^{3/2}}d\simeq 6\times10^{-8} m_8^{1/2}\,r_{16}^{-3/2}\,d, \label{eq:Delta_vK}
\end{equation} 
where $G$ is the gravitational constant and $m_8=M_{\rm BH}/10^8M_{\sun}$. For a slab to remain intact, the following condition should hold
\begin{equation}
\Delta v_{\rm K}<c_{\rm s},
\end{equation}
which is satisfied for
\begin{equation}
d<2\times10^{14}c_{\rm s,7}\,r_{16}^{3/2}\,m_8^{-1/2}\mbox{~cm},\label{eq:d_for_shear}
\end{equation}
where $c_{\rm s,7}=c_{\rm s}/10^7$~cm~s$^{-1}$, and $c_{\rm s,7}=1$ corresponds to $T\ga10^6$~K. For a cloud at $r_{\rm BLR}$ (Fig.~\ref{fig:struct_nT_17}), the outer layer where $n<10^8$~\cmt\ is expected to have $d>10^{14}$~cm, and thus be sheared. The colder inner layer, where $c_{\rm s,7}=0.1$, has $d<10^{13}$~cm which is only slightly smaller. The shear occurs on a timescale of
\begin{equation}
t_{\rm shear}\sim d/\Delta v_{\rm K}\simeq2\times10^7 r_{16}^{3/2}\,m_8^{-1/2}\mbox{~s}\simeq t_{\rm dyn}.
\end{equation}
If the BLR is composed of individual small clouds, than the clouds will likely be sheared on a dynamical timescale of a few. The likely steady state structure is a roughly an azimuthally symmetric gas distribution, as assumed in this study.

\subsection{Implications for observed trends of emission line properties} \label{sec:disc_obsr_trends}

The major trends in the emission line properties of AGN are the drop in EW of most lines with increasing $L$ (\citealt{baldwin77b}; and citations thereafter), and the so-called eigenvector 1 (EV1) set of correlation \citep{boroson_green92}. The Baldwin relation is commonly explained by a softening of the mean SED with increasing $L$ (e.g.\ \citealt{korista_etal98}), while the EV1 is commonly interpreted as a metallicity effect, where objects with high optical \FeII/\Hb, weak [\OIII]/\Hb\ and various other associated UV properties, such as the relative strength of the \NV~$\lambda$1240 line \citep{wills_etal99} have a high $Z$ \citep{shemmer_etal04}. RPC allows a more significant test of these suggestions, as \nH\ and $U$ are no longer free parameters.  Indeed, the RPC Cloudy model calculations presented above indicate that lines from ions of increasing ionization energy display a steeper dependence of the line EW on \aion, and since \aion\ and $L$ are likely related, it explains the steeper luminosity dependence of higher ionization lines \citep{dietrich_etal02}. Extreme EV1 objects, such as narrow line Seyfert 1 galaxies, generally have weaker \CIV~$\lambda$1549, stronger low ionization lines, such as \OI~$\lambda$1304, stronger N lines, such as \NIII]~$\lambda$1750, and stronger \SiIV~$\lambda$1397 (e.g.\ \citealt{wills_etal99}). These properties are indeed expected at a higher $Z$, as indicated by Fig.~\ref{fig:EW}. 

A very clear exception to the Baldwin relation is the \NV~$\lambda$1240 line, where the EW is essentially independent of $L$, despite its high ionization energy \citep{dietrich_etal02}. Fig.~\ref{fig:EW} demonstrates that this can be explained by a relation of $L$ and $Z$, where the steeper \aion\ with increasing $L$ is compensated by a rising $Z$ with $L$. The later relation is expected in a sample of luminous AGN (the $z\sim 3$ quasars used for measuring \NV~$\lambda$1240), which are inevitably close to their Eddington limit. Thus, high $L$ quasars are associated with a large black hole mass, which is associated with a massive host galaxy, which have a higher $Z$ \citep{hamann_ferland93, warner_etal03}. The higher than linear increase of the N abundance with $L$, increases the EW of various N lines, but for \NV~$\lambda$1240 this rise is almost exactly cancelled by the softer \aion, which decreases the EW of all other high ionization lines. This leads to an absence of a Baldwin relation for the \NV~$\lambda$1240 line, as was originally proposed by \citet{korista_etal98}. However, in these earlier studies $n$ and $U$ were considered as free parameters, while RPC indicates \aion\ and $Z$ are the only free parameters, and their combined effect nearly eliminates the Baldwin effect for the \NV~$\lambda$1240 line.

\subsection{The radial distribution of line emission}\label{sec:disc_rad_dist}

Figure~\ref{fig:EW} shows that the line emissivity drops sharply at $r>r_{\rm dust}$. The width of the transition region from dustless to dusty gas is expected to be broader than assumed here. The sublimation temperature depends on the grain composition, and the ambient pressure. In addition, the grain temperature depends on the grain size. The largest graphite grains ($0.25$~$\mu$m) sublimate at $r_{\rm dust}=0.2L_{\rm 46}^{1/2}$~pc (assuming negligible ambient pressure). The smallest graphite grains ($0.005$~$\mu$m) sublimate at $5.5r_{\rm dust}$, while the smallest silicate grains sublimate at $25r_{\rm dust}$ (\citealt{laor_draine93}, fig.~8 there). Thus, the dust sublimation region is actually extended, as the smallest silicate grains start to sublimate already at $5L_{\rm 46}^{1/2}$~pc, and complete sublimation occurs at $0.2L_{\rm 46}^{1/2}$~pc. These values also depend on the ambient pressure, where a higher pressure allows the grains to survive at a higher temperature. The grain size distribution, and the implied dust opacity therefore change with distance, and probably also with depth within the photoionized layer. For the sake of simplicity, we assume just a single $r_{\rm dust}$, but the drop in line emission beyond $r_{\rm dust}$ is expected to be more gradual.

Why does RPC predict a radial stratification in the line emission in the BLR (Fig.~\ref{fig:EW}), while RPC also predicts that $U\sim 0.1$ at all $\rb$? The reason is most likely a differential effect in the collisional suppression of lines of different ionization levels. The lower ionization lines come from $\rb\sim r_{\rm BLR}$, as they are collisionally suppressed at $\rb\ll r_{\rm BLR}$. In an RPC slab, a particular line emission (e.g.\ \CIV) originates in a layer with a local $U$ which is optimal in producing the emitting ion (C$^{3+}$). For slabs with smaller \rb\ (i.e.\ larger $n_\gamma$), this local $U$ is reached in layers with higher $n$. The lower ionization line emission (e.g.\ \CIV) is produced in layers with a relatively low $U$ ($\sim0.1$), which corresponds to very high $n$ for small \rb\ ($n\approx10^{13}$~\cmt\ for $\rb=0.03r_{\rm BLR}$; Fig.~\ref{fig:struct_all}, upper-left panel). But at these high $n$, the line becomes thermalized and produces little line emission (e.g.\ \citealt{ferland99}). The higher ionization lines (e.g.\ \NeVIII) are produced at a higher $U$ layer ($\sim10$), i.e.\ at lower $n$, which has a weaker effect in collisional suppression of the line emission. Another related effect is the depth and the column of the layer from which the various lines originate. Higher ionization lines come from a layer closer to the surface, with a lower column (since the flux of ionizing photons which can produce higher ionization ions is smaller). The lower column implies a smaller line optical depth, a higher line escape probability, and thus weaker collisional de-excitation. This effect is also seen in Cloudy uniform density slab models, where increasing $n$ at a fixed $U$ leads to a larger suppression of the lower ionization lines.

Figure~\ref{fig:EW} shows that if there is a uniform distribution of $\dd\Omega_{\rm BLR}/\dd r$, then the mean emission radius should drop with increasing line ionization. Thus, RM should show the largest delays for the \MgII\ and \CIII\ lines, followed by \Hb, \SiIII, \Lya, and \SiIV+\OIV. For \CIV\ the expected delay is about 1/3 of the low ionization lines delay, for \HeII\ and \OVI\ about 1/5, for \NV\ about 1/10, and for \NeVIII\ it is $<1/30$. In the best studied RM object, NGC~5548, the measured trend of delays with ionization is consistent with the above. The longest delays are of 40 and 22 days for \MgII\ and \CIII, 20 and 10 days for \Hb\ and \Lya, and 10, 2 and 2 days for  \CIV, \HeII\ and \NV\ (compilation in table 13.6 in \citealt{osterbrock_ferland06}). A similar effect is observed in ground based RM, where the \HeII~$\lambda 4686$ delay is a factor of $\sim 5-10$ shorter than the \Hb\ delay (e.g.\ \citealt{grier_etal13}). The similarity provides further support for the validity of RPC for the BLR, and further suggests that $\dd\Omega_{\rm BLR}/\dd r$ is roughly a constant in the BLR. How much does the BLR extend inwards? According to \citet{czerny_hryniewicz11} the BLR should terminate when the accretion disc atmosphere is too hot to support dust, which is expected to occur at a radius a factor of few smaller than $r_{\rm BLR}$. Thus, RM of the highest ionization lines, in particular the \NeVIII~$\lambda$774 line, can be used to probe the inner boundary of the BLR.

Another way to test the radial distribution of the emission of various lines is through their relative widths. \citet{hamann_etal98} indeed found that the \NeVIII~$\lambda 774$ line is significantly broader than other lines. Based on a handful of objects in table 2 there, we get a mean FWHM of 13,600~\kms\ for \NeVIII\, versus 6300~\kms\ for \OVI\ and 5200~\kms\ for \CIV, which for Keplerian motions suggest distance ratios of 7 for \NeVIII/\CIV\ and 1.5 for \OVI/\CIV, not far from the expected ratio based on Figure~\ref{fig:EW}. A similar effect can be seen in the higher quality COS spectra of lower $z$ AGN presented by \citet{shull_etal12}. Based on their composite spectrum (fig.~6 there), the estimated mean FWHM of \NeVIII~$\lambda 774$ is also 13,600~\kms\ (although the line is blended with nearby \NIV\ and \OIV\ lines; their contribution is expected to be negligible), versus a mean FWHM of 7,200~\kms\ for \OVI\ and 5,400~\kms\ for \CIV. Inspection of fig.~3 in \citet{shang_etal07} suggests that \HeII\ is typically broader than \CIV\ (see also \citealt{grier_etal13}), which is consistent with its smaller emitting radius (Fig.~\ref{fig:EW}). 

The \CIV\ line is often broader than \Hb, as expected from RPC (Fig.~\ref{fig:EW}), and consistent with the smaller size deduced from RM for the handful of objects with RM for the \CIV\ line. However, in objects where the \Hb\ FWHM$>4000$~\kms, one generally gets that the \Hb\ line is broader than \CIV\ (e.g.\ \citealt{baskin_laor05}). A specific striking case is Arp~102B, where the Balmer lines and the \MgII\ line are extremely broad and double peaked, while \CIV\ has a prominent single and relatively narrow peak \citep{halpern_etal96}. The low and high ionization lines in such objects may come from different regions \citep{collin_souffrin_etal88}. Such an effect may occur if the failed disc wind, which feeds the BLR, changes systematically with the global disc properties. In contrast, in 3C390.3, \CIV\ shows a double peaked structure, similar to the one seen in \Hb\ \citep{peterson_etal04}. The level of discrepancy between the high and low ionization line profiles is clearly an important and open question, which needs to be further explored.

\section{Conclusions}\label{sec:conclus}

AGN display similar emission line properties over a vast range of $10^8$ in luminosity. Here we point out that the ionizing $P_{\rm rad}$ sets $P_{\rm gas}$ inside a photoionized slab of gas. This RPC effect leads to a similar $U$ distribution within any photoionized static gas, regardless of its distance from the AGN, or the AGN luminosity. Since $r_{\rm BLR}$ is set by dust sublimation, either inside or above the accretion disc, which sets the $r_{\rm BLR}\propto L^{1/2}$ relation, the BLR is characterized by a universal $P_{\rm rad}\sim 0.3$~erg~cm$^{-3}$. The value of $P_{\rm gas}$ of an illuminated slab rises from the ambient pressure at the illuminated face of the slab, to $P_{\rm gas}=P_{\rm rad}$ deep inside the slab where all the radiation is absorbed, which implies $n\sim 10^{11}$~\cmt\ and $U\sim 0.1$ near the H ionization front. This leads to a universality of line ratios, as observed in AGN.

An RPC slab can be considered as a superposition of optically thin slabs with $U$ dropping from $>10^3$ near the surface, to $\le 0.1$ deep inside, where the last slab can be optically thick. An RPC slab thus produces lines from a larger range of ionization states, compared to a uniform density slab. The emitted spectrum includes lines from a large range of ionization states, and can reproduce the observed strength of most lines for $\Omega_{\rm BLR}\simeq 0.3$ (as suggested in earlier studies), including the \HeII\ line. There is no need to invoke a distribution of clouds with a range of $U$ values (which is in fact not possible for radially static clouds), as a specific range of $U$ is present in a single RPC slab.

RPC predicts a specific radial distribution of line emission, where the lower ionization lines peak near $r_{\rm dust}$, and the average emission radius moves inwards with increasing ionization state, as found in RM, and indicated by the larger FWHM of the higher ionization lines, such as \NeVIII, \OVI\ and \HeII. 

The free parameters which affect the RPC solution at a given $r$ are $\aion$ and $Z$. The derived trends of the line strength with $\aion$ and $Z$ are similar to those derived in uniform density models, where the recombination lines provide a good measure of $\aion$, while the \NV\ line and the weaker metal lines are sensitive to $Z$. However, RPC provides a unique solution for the integrated line emission, for a given $\dd\Omega_{\rm BLR}/\dd r$. This solution can be used to test the validity of possible mechanisms that may set $\dd\Omega_{\rm BLR}/\dd r$.

An apparent inconsistency with RPC appears in objects with very broad Balmer lines, which generally show narrower \CIV\ lines. This may indicate some fundamental change in the BLR structure for some specific global accretion parameters, and needs to be further explored.

\section*{Acknowledgments}
We thank the referee, A.\ Marconi, for valuable comments. We acknowledge fruitful discussion with H.\ Netzer. We thank G.\ Ferland for developing Cloudy and making it publicly available. Without his work, this study would not evolve beyond the analytic approximations. This research has made use of NASA's Astrophysics Data System Bibliographic Services.

\appendix
\section{Analytic treatment of an RPC slab with  $\bmath{\lowercase{d}\sim \lowercase{\ri}}$}\label{sec:large_d}
We present below an analytic approximation for the slab $n(r)$ and total depth $d$, when $d\sim\ri$ and the geometrical dilution of continuum radiation should be accounted for. We follow similar considerations to those of Section~\ref{sec:th_qual}, and assume an equilibrium between gas and radiation pressure. We also assume that the gas is at a constant $T\approx \Tcomp$, and that the continuum is attenuated by electron scattering only. These assumptions yield
\begin{equation}
 k\Tcomp\textrm{d}n_{\rm tot}=\frac{\sigma_{\rm es}}{c}\frac{L}{4\pi(\ri+d^\prime)^2}n_e\textrm{d}d^\prime,\label{eq:appA_P_equi}
\end{equation}
where $n_{\rm tot}$ is the number density that is summed over all gas constituents, and $d^\prime$ is the distance from \ri\ (i.e.\ $d^\prime=r-\ri$). For a fully ionized gas with $Z=Z_{\sun}$, $n_e=1.15\nH$ and $n_{\rm tot}\approx2.15\nH$. Solution of eq.~\ref{eq:appA_P_equi} with the boundary condition of $n_{\rm H}=n_{{\rm H,i}}$ at $d^\prime=0$ implies
\begin{equation}
 \nH=\nHi\exp{A\left(\frac{1}{\ri}-\frac{1}{\ri+d^\prime}\right)},\label{eq:appA_nH}
\end{equation}
\begin{equation}
  A\equiv\frac{1.15}{2.15}\frac{\sigma_{\rm es}}{c}\frac{L}{4\pi}\frac{1}{k\Tcomp}.
\end{equation}
This analytic approximation of \nH\ is presented in Fig.~\ref{fig:struct_nT_17} (middle-right panel). Note that for $d\ll \ri$, 
\begin{equation}
 \nH=\nHi\exp{A d^\prime/\ri^2}, 
\end{equation}
and a typical length-scale can be defined 
\begin{equation}
 l_{\rm pr}^{-1}\equiv A/\ri^2,
\end{equation}
 which is equivalent to eq.~\ref{eq:r_pr}.

Noting that the slab $d$ is approximately the depth of the completely ionized region (e.g.\ Figs~\ref{fig:struct_nT_15} and \ref{fig:struct_nT_17}), one can derive an analytic approximation of $d$. We define $U_{\rm l}$ as the ionization parameter at which $T<\Tcomp$, and attenuation of the continuum by ionic transitions becomes important ($U_{\rm l}\approx10^{3}$, e.g.\ \citealt{hamann97}). As noted above (Section~\ref{sec:results_slab_strc}), the optical depth of the hot layer of the slab, where $T\approx\Tcomp$ and $U>U_{\rm l}$, is small ($\tau\sim0.01$). This implies that $U_{\rm l}$ can be approximated by
\begin{equation}
 U_{\rm l}\approx\frac{L_{\rm ion}}{4\pi(\ri+d)^2\hnu_{\rm ion}c}\frac{1}{\nH}.
\end{equation}
Inserting \nH\ from eq.~\ref{eq:appA_nH} and $L_{\rm ion}=f_{\rm ion}L$ yields
\begin{equation}
 \exp{A\left(\frac{1}{\ri}-\frac{1}{\ri+d}\right)}=\frac{B}{(\ri+d)^2},
\end{equation}
\begin{equation}
 B\equiv\frac{f_{\rm ion} L}{4\pi\hnu_{\rm ion}c}\frac{1}{\nH_{\rm i}U_{\rm l}}.
\end{equation}
Defining $D\equiv \ri+d$ and $W\equiv A/2D$ yields
\begin{equation}
 W\textrm{e}^W=\frac{A}{2\sqrt{B}}\exp{\frac{A}{2\ri}},
\end{equation}
where $W$ is the Lambert W function, which can be solved for numerically.

\section{Comparison to models which include \Pl}\label{sec:comp_Pline}

\begin{figure}
 \includegraphics[width=84mm]{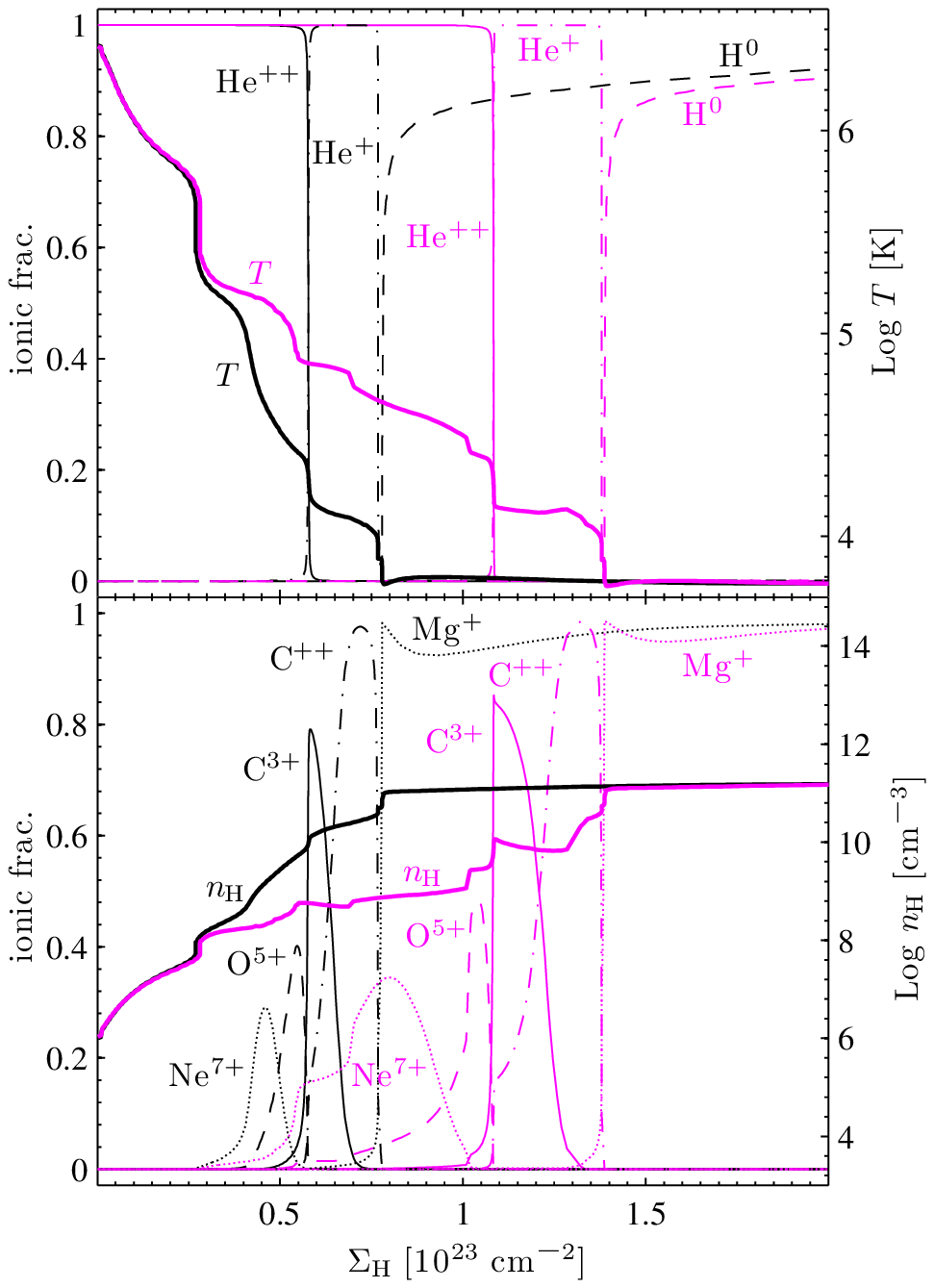}
 \caption{Same as Fig.~\ref{fig:struct_all}, but also presents an RPC model which includes \Pl\ (magenta lines). The models adopt $\nHi=10^6$~\cmt\ and $\rb=r_{\rm BLR}$. The ionization structure of the model with \Pl\ is shifted in $\Sigma_{\rm H}$ by $\simeq+0.6\times10^{23}$~\cmmt\ compared to the model without \Pl.}\label{fig:struct_17_pline}
\end{figure}

Figures \ref{fig:struct_nT_15}, \ref{fig:struct_nT_17} and \ref{fig:struct_17_pline} compare the RPC slab structure between models with and without \Pl. The slab structure versus depth of $\nHi=10^6$ and $10^8$~\cmt\ models that include \Pl\ is presented for $\rb=0.03r_{\rm BLR}$ and $r_{\rm BLR}$ (Figs~\ref{fig:struct_nT_15} and \ref{fig:struct_nT_17}, respectively). Similarly to models without \Pl, when the slab depth is measured from \rb, the two different \nHi\ models with \Pl\ overlap remarkably well for both values of \rb. Models with and without \Pl\ have an overall similar structure (for a given \nHi; Figs~\ref{fig:struct_nT_15} and \ref{fig:struct_nT_17}, left panels), and reach the same \nHb. The two types of models differ at depths that correspond to $10^6\ga T\ga10^4$~K (Figs~\ref{fig:struct_nT_15} and \ref{fig:struct_nT_17}, right panels). At these depths, models with \Pl\ have lower $T$ and larger $n$ by up to 0.5 and 1~dex, respectively.

Figure~\ref{fig:struct_17_pline} presents the slab structure versus \NH\ for $\nHi=10^6$~\cmt\ and $\rb=r_{\rm BLR}$ model with and without \Pl. There are two main differences between the two models. First, the model with \Pl\ has lower \nH\ and higher $T$ for $0.3 <\NH/10^{23}<1.4$~\cmmt. The difference in $nT$ (not shown) implies that \Pl\ can be as high as $2P_{\rm gas}$ in regions where the difference is maximal. Secondly, the whole ionization structure of the model with \Pl\ is shifted by $\approx+0.6\times10^{23}$~\cmmt\ relative to the model without \Pl. These two differences can be explained as follows. For $\NH\la0.3\times10^{23}$~\cmmt, \Pl\ is negligible as there is little resonance line emission for $T\ga10^{5.5}$~K. For larger \NH\ (lower $T$), \Pl\ is non-negligible and $P_{\rm gas}$ is reduced. Lower $P_{\rm gas}$ yields smaller \nH\ compared to the model without \Pl, which implies larger $T$, since cooling becomes less efficient.\footnote{Lower $P_{\rm gas}$ values cannot be achieved by a process which reduces $T$, since for $T\la10^{5.5}$~K, lower $T$ implies a more efficient cooling, and this process is unstable.} Larger values of $T$ yield lower values of the ionic recombination coefficient (e.g.\ \citealt{verner_ferland96}), less absorption of ionizing radiation by $\dd\NH$, and a larger \NH\ which is needed to reach a particular ionization sate (e.g.\ the H ionization front), as apparent from Fig.~\ref{fig:struct_17_pline}.

Figure~\ref{fig:EW} presents the effect of \Pl\ on line emission for models with $Z=Z_{\sun}$ and $\aion=-1.6$. Including \Pl\ has little effect on emission of most of the lines presented in Fig.~\ref{fig:EW}. It has some effect (a factor $\la2$) on \NeVIII, \NV, \OIV] and \CIV, increasing the \NeVIII\ emission, and decreasing the emission of the other three lines, due to increase in $U$. However, the low ionization lines (with an ionization potential $<1$~Ryd) are not affected, since the inner ionization structure is not affected by \Pl. The recombination dominated lines (\HeII, \Lya\ and \Hb) are also not affected, as expected since these lines just count the number of incident ionizing photons.

It should be mentioned in passing that the magnitude of \Pl\ is likely to be smaller in BLR than predicted by the photoionization models which are presented here. While our models include only thermal line-broadening, the BLR gas is probably affected by a mechanism that produces a more significant line-broadening. The mechanism might be either a significant supersonic turbulent velocity \citep{bottorff_etal00}, a large scale flow, e.g.\ as expected in a failed disc wind, or the shear in the Keplerian velocity which becomes significant if the condition $d\ll\ri$ does not hold (Appendix~\ref{sec:large_d}). This broadening mechanism will strongly reduce the trapping of resonance lines, and thus reduce \Pl\ and its effects on gas emission. Addressing the effects of the broadening mechanism on the structure of RPC BLR is beyond the scope of this paper.

\bsp
\label{lastpage}

\begin{thebibliography}{99}
\bibitem[\protect\citeauthoryear{Arav et al.}{1998}]{arav_etal98} Arav N., Barlow T.~A., Laor A., Sargent W.~L.~W., Blandford R.~D., 1998, MNRAS, 297, 990 
\bibitem[\protect\citeauthoryear{Baldwin}{1977a}]{baldwin77a} Baldwin J.~A., 1977a, MNRAS, 178, 67P
\bibitem[\protect\citeauthoryear{Baldwin}{1977b}]{baldwin77b} Baldwin J.~A., 1977b, ApJ, 214, 679
\bibitem[\protect\citeauthoryear{Baldwin et al.}{1995}]{baldwin_etal95} Baldwin J., Ferland G., Korista K., Verner D., 1995, ApJ, 455, L119 
\bibitem[\protect\citeauthoryear{Baldwin et  al.}{1996}]{baldwin_etal96} Baldwin J.~A., et al., 1996, ApJ, 461, 664
\bibitem[\protect\citeauthoryear{Baldwin et al.}{2004}]{baldwin_etal04} Baldwin J.~A., Ferland G.~J., Korista K.~T., Hamann F., LaCluyz{\'e} A., 2004, ApJ, 615, 610
\bibitem[\protect\citeauthoryear{Baskin \& Laor}{2005}]{baskin_laor05} Baskin A., Laor A., 2005, MNRAS, 356, 1029 
\bibitem[\protect\citeauthoryear{Begelman, de Kool  \& Sikora}{Begelman et al.}{1991}]{begelman_etal91} Begelman M., de Kool M., Sikora M., 1991, ApJ, 382, 416 
\bibitem[\protect\citeauthoryear{Binette et al.}{1997}]{binette_etal97} Binette L., Wilson A.~S., Raga A., Storchi-Bergmann T., 1997, A\&A, 327, 909 
\bibitem[\protect\citeauthoryear{Blumenthal  \& Mathews}{1975}]{blumenthal_mathews75} Blumenthal G.~R., Mathews W.~G., 1975, ApJ, 198, 517 
\bibitem[\protect\citeauthoryear{Boroson \& Green}{1992}]{boroson_green92} Boroson T.\ A., Green R.\ F., 1992, ApJS, 80, 109
\bibitem[\protect\citeauthoryear{Bottorff et al.}{2000}]{bottorff_etal00} Bottorff M., Ferland G., Baldwin J., Korista K., 2000, ApJ, 542, 644 
\bibitem[\protect\citeauthoryear{Brandt, Laor \& Wills}{Brandt et al.}{2000}]{brandt_etal00} Brandt W.\ N., Laor A., Wills B.\ J., 2000, ApJ, 528, 637
\bibitem[\protect\citeauthoryear{Capriotti, Foltz \& Byard}{Capriotti et al.}{1981}]{capriotti_etal81} Capriotti E., Foltz C., Byard P., 1981, ApJ, 245, 396 
\bibitem[\protect\citeauthoryear{Chelouche \& Netzer}{2001}]{chelouche_netzer01} Chelouche D., Netzer H., 2001, MNRAS, 326, 916
\bibitem[\protect\citeauthoryear{Collin-Souffrin et al.}{1988}]{collin_souffrin_etal88} Collin-Souffrin S., Dyson J.~E., McDowell J.~C., Perry J.~J., 1988, MNRAS, 232, 539 
\bibitem[\protect\citeauthoryear{Croom et al.}{2002}]{croom_etal02} Croom S.~M., et al., 2002, MNRAS, 337, 275 
\bibitem[\protect\citeauthoryear{Czerny \& Hryniewicz}{2011}]{czerny_hryniewicz11} Czerny B., Hryniewicz K., 2011, A\&A, 525, L8
\bibitem[\protect\citeauthoryear{Czerny et al.}{2009}]{czerny_etal09} Czerny B., Chevallier L., Gon{\c c}alves A.\ C., R{\'o}{\.z}a{\'n}ska A., Dumont, A.-M.\ 2009, A\&A, 499, 349 
\bibitem[\protect\citeauthoryear{Davidson}{1972}]{davidson72} Davidson K., 1972, ApJ, 171, 213 
\bibitem[\protect\citeauthoryear{Davidson \& Netzer}{1979}]{davidson_netzer79} Davidson K., Netzer H., 1979, RvMP, 51, 715
\bibitem[\protect\citeauthoryear{Dietrich et al.}{2002}]{dietrich_etal02} Dietrich M., Hamann F., Shields J.~C., Constantin A., Vestergaard M., Chaffee F., Foltz C.~B., Junkkarinen V.~T., 2002, ApJ, 581, 912 
\bibitem[\protect\citeauthoryear{Dopita et al.}{2002}]{dopita_etal02} Dopita M.~A., Groves B.~A., Sutherland R.~S., Binette L., Cecil G., 2002, ApJ, 572, 753 
\bibitem[\protect\citeauthoryear{Draine}{2011a}]{draine11a} Draine B.~T., 2011a, ApJ, 732, 100 
\bibitem[\protect\citeauthoryear{Draine}{2011b}]{draine11b} Draine B.~T., 2011b, Physics of the Interstellar and Intergalactic Medium. Princeton Univ. Press, Princeton, NJ
\bibitem[\protect\citeauthoryear{Emmering, Blandford \& Shlosman}{Emmering et al.}{1992}]{emmering_etal92} Emmering R.~T., Blandford R.~D., Shlosman I., 1992, ApJ, 385, 460 
\bibitem[\protect\citeauthoryear{Ferland}{1999}]{ferland99} Ferland G., 1999, in Ferland G., Baldwin J., eds, ASP Conf.\ Ser.\ Vol.\ 162, Quasars and Cosmology. Astron.\ Soc.\ Pac., San Francisco, p.~147
\bibitem[\protect\citeauthoryear{Ferland et al.}{1998}]{ferland_etal98} Ferland G.\ J., Korista K.\ T., Verner D.\ A., Ferguson J.\ W., Kingdon J.\ B., Verner E.\ M., 1998, PASP, 110, 761
\bibitem[\protect\citeauthoryear{Gon{\c c}alves et al.}{2007}]{goncalves_etal07} Gon{\c c}alves A.\ C., Collin S., Dumont A.-M., Chevallier L.\ 2007, A\&A, 465, 9
\bibitem[\protect\citeauthoryear{Grier et al.}{2013}]{grier_etal13} Grier C.~J., et al., 2013, ApJ, 764, 47 
\bibitem[\protect\citeauthoryear{Groves, Dopita \& Sutherland}{Groves et al.}{2004}]{groves_etal04} Groves B.\ A., Dopita M.\ A., Sutherland R.\ S., 2004, ApJS, 153, 9
\bibitem[\protect\citeauthoryear{Halpern et al.}{1996}]{halpern_etal96} Halpern J.~P., Eracleous M., Filippenko A.~V., Chen K., 1996, ApJ, 464, 704 
\bibitem[\protect\citeauthoryear{Hamann}{1997}]{hamann97} Hamann F., 1997, ApJS, 109, 279
\bibitem[\protect\citeauthoryear{Hamann \& Ferland}{1993}]{hamann_ferland93} Hamann F., Ferland G., 1993, ApJ, 418, 11 
\bibitem[\protect\citeauthoryear{Hamann \& Ferland}{1999}]{hamann_ferland99} Hamann F., Ferland G., 1999, ARA\&A, 37, 487 
\bibitem[\protect\citeauthoryear{Hamann et al.}{1998}]{hamann_etal98} Hamann F., Cohen R.~D., Shields J.~C., Burbidge E.~M., Junkkarinen V., Crenshaw D.~M., 1998, ApJ, 496, 761
\bibitem[\protect\citeauthoryear{Kaspi et al.}{2005}]{kaspi_etal05} Kaspi S., Maoz D., Netzer H., Peterson B.\ M., Vestergaard M., Jannuzi B.\ T., 2005, ApJ, 629, 61
\bibitem[\protect\citeauthoryear{Kaspi et al.}{2007}]{kaspi_etal07}  Kaspi S., Brandt W.~N., Maoz D., Netzer H., Schneider D.~P., Shemmer O., 2007, ApJ, 659, 997
\bibitem[\protect\citeauthoryear{Korista et al.}{1997}]{korista_etal97} Korista K., Baldwin J., Ferland G., Verner D., 1997, ApJS, 108, 401 
\bibitem[\protect\citeauthoryear{Korista, Baldwin, \& Ferland}{Korista et al.}{1998}]{korista_etal98} Korista K., Baldwin J., Ferland G., 1998, ApJ, 507, 24 
\bibitem[\protect\citeauthoryear{Kraemer et al.}{1999}]{kraemer_etal99} Kraemer S.~B., Ho L.~C., Crenshaw D.~M., Shields J.~C., Filippenko A.~V., 1999, ApJ, 520, 564 
\bibitem[\protect\citeauthoryear{Krolik}{1988}]{krolik88} Krolik J.~H., 1988, ApJ, 325, 148 
\bibitem[\protect\citeauthoryear{Krolik}{1999}]{krolik99} Krolik J.~H., 1999, Active  galactic nuclei: from the central black hole to the galactic environment. Princeton Univ. Press, Princeton, NJ 
\bibitem[\protect\citeauthoryear{Krolik \& Kriss}{2001}]{krolik_kriss01} Krolik J.~H., Kriss G.~A., 2001, ApJ, 561, 684 
\bibitem[\protect\citeauthoryear{Laor}{1998}]{laor98} Laor A., 1998, ApJ, 505, L83 
\bibitem[\protect\citeauthoryear{Laor \& Draine}{1993}]{laor_draine93} Laor A., Draine B.\ T., 1993, ApJ, 402, 441
\bibitem[\protect\citeauthoryear{Laor et al.}{2006}]{laor_etal06} Laor A., Barth A.~J., Ho L.~C., Filippenko A.~V., 2006, ApJ, 636, 83 
\bibitem[\protect\citeauthoryear{Magorrian et al.}{1998}]{magorrian_etal98} Magorrian J., et al., 1998, AJ, 115, 2285
\bibitem[\protect\citeauthoryear{Maiolino et al.}{2001}]{maiolino_elal01} Maiolino R., Salvati M., Marconi A., Antonucci R.~R.~J., 2001, A\&A, 375, 25 
\bibitem[\protect\citeauthoryear{Maoz et al.}{1991}]{maoz_etal91} Maoz D., et al., 1991, ApJ, 367, 493
\bibitem[\protect\citeauthoryear{Marconi et al.}{2008}]{marconi_etal08} Marconi A., Axon D.~J., Maiolino R., Nagao T., Pastorini G., Pietrini P., Robinson A., Torricelli G., 2008, ApJ, 678, 693
\bibitem[\protect\citeauthoryear{Marconi et al.}{2009}]{marconi_etal09} Marconi A., Axon D.~J., Maiolino R., Nagao T., Pietrini P., Risaliti G., Robinson A., Torricelli G., 2009, ApJ, 698, L103
\bibitem[\protect\citeauthoryear{Mathews}{1982}]{mathews82} Mathews W.~G., 1982, ApJ, 252, 39 
\bibitem[\protect\citeauthoryear{Mathews}{1986}]{mathews86} Mathews W.~G., 1986, ApJ, 305, 187 
\bibitem[\protect\citeauthoryear{Mathews \& Blumenthal}{1977}]{mathews_blumenthal77} Mathews W.~G., Blumenthal G.~R., 1977, ApJ, 214, 10 
\bibitem[\protect\citeauthoryear{Mathews  \& Doane}{1990}]{mathews_doane90} Mathews W.~G., Doane J.~S., 1990, ApJ, 352, 423 
\bibitem[\protect\citeauthoryear{Mathews \& Ferland}{1987}]{mathews_ferland87} Mathews W.~G., Ferland G.~J., 1987, ApJ, 323, 456 
\bibitem[\protect\citeauthoryear{Netzer}{1976}]{netzer76} Netzer H., 1976, MNRAS, 177, 473
\bibitem[\protect\citeauthoryear{Netzer \& Laor}{1993}]{netzer_laor93} Netzer H., Laor A., 1993, ApJ, 404, L51
\bibitem[\protect\citeauthoryear{Netzer et al.}{1995}]{netzer_etal95} Netzer H., Brotherton M.~S., Wills B.~J., Han M., Wills D., Baldwin J.~A., Ferland G.~J., Browne I.~W.~A., 1995, ApJ, 448, 27 
\bibitem[\protect\citeauthoryear{Osterbrock \& Ferland}{2006}]{osterbrock_ferland06} Osterbrock D.~E., Ferland, G.~J., 2006, Astrophysics of gaseous nebulae and active galactic nuclei, 2nd.~ed.. Univ.\ Science Books, Sausalito, CA
\bibitem[\protect\citeauthoryear{Perry \& Dyson}{1985}]{perry_dyson85} Perry J.~J., Dyson J.~E., 1985, MNRAS, 213, 665 
\bibitem[\protect\citeauthoryear{Peterson et al.}{2004}]{peterson_etal04} Peterson B.~M., et al., 2004, ApJ, 613, 682 
\bibitem[\protect\citeauthoryear{Rees}{1987}]{rees87} Rees M.~J., 1987, MNRAS, 228, 47P 
\bibitem[\protect\citeauthoryear{Rees, Netzer \& Ferland}{Rees et al.}{1989}]{rees_etal89} Rees M.~J., Netzer H., Ferland G.~J., 1989, ApJ, 347, 640 
\bibitem[\protect\citeauthoryear{Risaliti et al.}{2011}]{risaliti_etal11} Risaliti G., Nardini E., Salvati M., Elvis M., Fabbiano G., Maiolino R., Pietrini P., Torricelli-Ciamponi G., 2011, MNRAS, 410, 1027
\bibitem[\protect\citeauthoryear{R{\'o}{\.z}a{\'n}ska et al.}{2006}]{rozanska_etal06} R{\'o}{\.z}a{\'n}ska A., Goosmann R., Dumont A.-M., Czerny B., 2006, A\&A, 452, 1 
\bibitem[\protect\citeauthoryear{Ruff et al.}{2012}]{ruff_etal12} Ruff A.~J., Floyd D.~J.~E., Webster R.~L., Korista K.~T., Landt H., 2012, ApJ, 754, 18  
\bibitem[\protect\citeauthoryear{Shang et al.}{2007}]{shang_etal07} Shang Z., Wills B.~J., Wills D., Brotherton M.~S., 2007, AJ, 134, 294 
\bibitem[\protect\citeauthoryear{Shemmer et al.}{2004}]{shemmer_etal04} Shemmer O., Netzer H., Maiolino R., Oliva E., Croom S., Corbett E., di Fabrizio L., 2004, ApJ, 614, 547 
\bibitem[\protect\citeauthoryear{Shen et al.}{2011}]{shen_etal11} Shen Y., et al., 2011, ApJS, 194, 45
\bibitem[\protect\citeauthoryear{Shields}{1978}]{shields78} Shields G.\ A., 1978, in Wolfe A.~M., ed., Pittsburgh Conference on BL Lac Objects. Univ.\ of Pittsburgh, Pittsburgh, PA, p.~257
\bibitem[\protect\citeauthoryear{Shull, Stevans \& Danforth}{Shull et al.}{2012}]{shull_etal12} Shull J.~M., Stevans M., Danforth C.~W., 2012, ApJ, 752, 162 
\bibitem[\protect\citeauthoryear{Steffen et al.}{2006}]{steffen_etal06} Steffen A.~T., Strateva I., Brandt W.~N., Alexander D.~M., Koekemoer A.~M., Lehmer B.~D., Schneider D.~P., Vignali C., 2006, AJ, 131, 2826 
\bibitem[\protect\citeauthoryear{Stern, Laor \& Baskin}{Stern et al.}{2013}]{stern_etal13} Stern J., Laor A., Baskin A., in prep. (Paper I)
\bibitem[\protect\citeauthoryear{Steenbrugge et al.}{2005}]{steenbrugge_etal05} Steenbrugge K.~C., et al., 2005, A\&A, 434, 569 
\bibitem[\protect\citeauthoryear{Suganuma et al.}{2006}]{suganuma_etal06} Suganuma M., et al., 2006, ApJ, 639, 46 
\bibitem[\protect\citeauthoryear{Telfer et al.}{2002}]{telfer_etal02} Telfer R.\ C., Zheng W., Kriss G.\ A., Davidsen A.\ F., 2002, ApJ, 565, 773
\bibitem[\protect\citeauthoryear{Verner \& Ferland}{1996}]{verner_ferland96} Verner D.~A., Ferland G.~J., 1996, ApJS, 103, 467 
\bibitem[\protect\citeauthoryear{Warner, Hamann \& Dietrich}{Warner et al.}{2003}]{warner_etal03} Warner C., Hamann F., Dietrich M., 2003, ApJ, 596, 72 
\bibitem[\protect\citeauthoryear{Wills et al.}{1999}]{wills_etal99} Wills B.~J., Laor A., Brotherton M.~S., Wills D., Wilkes B.~J., Ferland G.~J., Shang Z., 1999, ApJ, 515, L53 
\bibitem[\protect\citeauthoryear{Yeh et al.}{2013}]{yeh_etal2013} Yeh S.~C.~C., Verdolini S., Krumholz M.~R., Matzner C.~D., Tielens A.~G.~G.~M., 2013, ApJ, 769, 11 
\end{thebibliography}
\end{document}